\documentclass[onecolumn,usenatbib,usegraphicx]{mn2e}

\newcommand{\sig}{\sigma_{_{\rm T}}}
\newcommand{\oc}{\bar n}
\newcommand{\begeq}{\begin{equation}}
\newcommand{\fineq}{\end{equation}}
\newcommand{\xmin}{x_{\rm min}}
\newcommand{\xmax}{x_{\rm max}}
\newcommand{\green}{f_{_{\rm G}}}
\newcommand{\greenoc}{\bar n_{_{\rm G}}}
\newcommand{\greendens}{\eta}
\newcommand{\greenmom}{I^{\rm G}}
\newcommand{\wien}{f_{_{\rm W}}}
\newcommand{\Tic}{T_{_{\rm IC}}}
\newcommand{\mnras}{MNRAS}
\newcommand{\apj}{ApJ}
\newcommand{\apss}{Ap\&SS}
\newcommand{\aap}{A\&A}
\newcommand{\greensoft}{f_{_{\rm soft}}}

\title[Green's function for thermal Comptonization]
{Exact solution for the Green's function describing
time-dependent thermal Comptonization}
\author[P. A. Becker]{Peter A. Becker\thanks{E-mail:
pbecker@gmu.edu}\\
Center for Earth Observing and Space Research, George Mason
University, Fairfax, VA 22030-4444, USA}

\begin{document}

\date{Submitted 2002 December 26.
      Received ;
      in original form}

\pagerange{\pageref{firstpage}--\pageref{lastpage}} \pubyear{2002}

\maketitle

\label{firstpage}

\begin{abstract}
We obtain an exact, closed-form expression for the time-dependent
Green's function solution to the Kompaneets equation. The result, which
is expressed as the integral of a product of two Whittaker functions,
describes the evolution in energy space of a photon distribution that is
initially monoenergetic. Effects of spatial transport within a
homogeneous scattering cloud are also included within the formalism. The
Kompaneets equation that we solve includes both the recoil and energy
diffusion terms, and therefore our solution for the Green's function
approaches the Wien spectrum at large times. This was not the case with
earlier analytical solutions that neglected the recoil term and were
therefore applicable only in the soft-photon limit. We show that the
Green's function can be used to generate all of the previously known
steady-state and time-dependent solutions to the Kompaneets equation.
The new solution allows the direct determination of the spectrum,
without the need to numerically solve the partial differential equation.
It is therefore much more convenient for data analysis purposes. Based
upon the Green's function, we derive a new, exact solution for the
variation of the inverse-Compton temperature of an initially
monoenergetic photon distribution. Furthermore, we also obtain a new
time-dependent solution for the photon distribution resulting from the
reprocessing of an optically thin bremsstrahlung initial spectrum with a
low-energy cutoff. Unlike the previously known solution for
bremsstrahlung injection, the new solution possesses a finite photon
number density, and therefore it displays proper equilibration to a Wien
spectrum at large times. The relevance of our results for the
interpretation of emission from variable X-ray sources is discussed,
with particular attention to the production of hard X-ray time lags, and
the Compton broadening of narrow features such as iron lines.
\end{abstract}


\section{INTRODUCTION}

In hot, radiation-dominated plasmas, the primary interaction between
photons and electrons occurs via Compton scattering. This process
consequently plays a central role in the formation of X-ray spectra in a
variety of sources, including active galaxies, low-mass X-ray binaries,
and the early universe. The repeated scattering of radiation off free
electrons (Comptonization) can have a profound effect on both the X-ray
spectrum and the Fourier structure of the observed emission in the time
domain. The effect of multiple Compton scattering depends on the
velocity distribution of the electrons; random and systematic motions of
the electrons produce ``thermal'' and ``bulk'' Comptonization,
respectively. Thermal Comptonization can explain the formation of the
power-law tails observed in the X-ray spectra from active galaxies and
low-mass X-ray binaries, as well as the development of time lags between
the hard and soft photons (Shapiro, Lightman, \& Eardley 1976; Sunyaev
\& Titarchuk 1980; van der Klis et al. 1987; Stollman et al. 1987). This
process also governs the evolution of the cosmic background radiation
before recombination (Sunyaev \& Zeldovich 1970; Zeldovich \& Sunyaev
1969; Illarionov \& Sunyaev 1975a, 1975b), and it is a necessary
ingredient in models for the production of the modified blackbody
continuum emission observed during X-ray bursts (Sunyaev \& Titarchuk
1980; Titarchuk 1988). Thermal Comptonization also seems to contribute
to the broadening of the K$\alpha$ iron lines observed in a number of
sources (Reynolds \& Wilms 2000; Wang, Zhou, \& Wang 1999; Ross, Fabian,
\& Young 1999). In addition to the thermal process, bulk Comptonization
can also influence the spectral formation when the electrons are
converging on average, as for example in a radiation-dominated shock or
a quasi-spherical accretion flow (Payne \& Blandford 1981; Lyubarskii \&
Sunyaev 1982; Becker 1988; Colpi 1988; Titarchuk, Mastichiadis, \&
Kylafis 1997; Titarchuk \& Zannias 1998; Laurent \& Titarchuk 2001;
Titarchuk \& Shrader 2002).

The effect of thermal Comptonization on the photon distribution is
described by the well-known Kompaneets partial differential equation,
published in 1957. This equation, based on a Fokker-Planck formalism,
describes the effect of multiple Compton scattering on the photon
distribution when the electrons are moving nonrelativistically and the
average fractional energy change per scattering is small. The Kompaneets
equation has provided the foundation for most of the theoretical studies
of thermal Comptonization in astrophysical environments, which have
focused primarily on the steady-state reprocessing of monoenergectic
radiation injected continuously into a nonrelativistic plasma with
time-independent properties (e.g., Katz 1976; Sunyaev \& Titarchuk
1980). These studies have provided useful insight into the formation of
the ubiquitous power-law spectra observed in the quiescent emission from
active galactic nuclei (AGNs) and galactic black-hole candidates.
Titarchuk \& Lyubarskij (1995) generalized these results by solved the
stationary form of the Boltzmann kinetic equation to demonstrate that
power-law spectra can also result from Comptonization in a cloud of
relativistic electrons. Despite the success of the steady-state models
in helping us to understand the formation of the quiescent X-ray
spectra, they cannot be used to follow the spectral and temporal
evolution of rapid transients. This is a concern because rapid
variability is a characteristic feature of many classes of X-ray
sources. Indeed, the rapid variability itself contains detailed
information about the geometry and spatial structure of the inner region
of the accretion flow. Analysis of the ``Compton reverberations''
associated with the impulsive injection of soft photons can provide
perhaps the most direct probe of the inner region (Reynolds et al. 1999;
Ulrich 2000). Specific examples of rapid variability include the strong
X-ray flares observed from the Seyfert galaxies NGC~4151 (Tananbaum et
al. 1978; Lawrence 1980) and NGC~6814 (K\"onig et al. 1997; Mittaz \&
Branduardi-Raymont 1989) in which the intensity increases by a factor of
5-10 in $\la 1000\,$s. Many other active galaxies also display
significant variability (see, e.g., Iwasawa et al. 2000; Merloni \&
Fabian 2001), and flickering and bimodal spectral behavior on timescales
of seconds involving sustantial fractions of the X-ray flux have been
observed from Cyg X-1 (Tanaka 1989; Zdziarski et al. 2002), GX~339-4
(Kitamoto 1989; Miyamoto et al. 1991), and other galactic black-hole
candidates (Mereghetti 1993; Smith, Heindl, \& Swank 2002).

Observations of rapid variability during strong X-ray transients in AGNs
and galactic black-hole candidates suggest that a time-dependent model
for the spectral formation may be required in order to understand the
physical processes occurring in the source plasma. Surprisingly, the
time-dependent Green's function associated with the Kompaneets equation
has never been obtained in closed form. Time-dependent analytical
solutions previously discussed in the literature have focused on certain
special cases, or have adopted simplifying approximations. For example,
Chapline and Stevens (1973) examined the Comptonization of a
bremsstrahlung initial spectrum in a plasma with steady properties.
However, the time-dependent analytical solution they obtain is
unphysical in the sense that full equilibration to a Wien spectrum never
occurs because the initial spectrum contains an infinite number of
low-energy photons. Zeldovich \& Sunyaev (1969) and Payne (1980) derived
the Green's function for time-dependent thermal Comptonization under the
assumption that $\epsilon \ll kT_e$, where $T_e$ is the electron
temperature and $\epsilon$ is the photon energy. Under this restriction,
the effect of electron recoil can be ignored, and the transport equation
admits a relatively simple solution for the Green's function. However,
this ``soft-photon'' Green's function is not valid when photons remain
in the plasma long enough to upscatter to energies comparable to the
electron thermal energy, which cannot be ruled out in the observed
transients. Futher discussion of this point is provided in \S~6.1.

More general results have been obtained using numerical simulations
based on a modified form of the Kompaneets equation that includes an
additional term describing spatial diffusion. For example, B\"ottcher \&
Liang (1998) simulated numerically the Compton upscattering of flares of
soft radiation in clouds with various geometries, and Malzac \& Jourdain
(2000) utilized a Monte-Carlo code to model time-dependent Comptonizing
flares in a corona with a self-consistently determined temperature. The
utilization of partial differential equation solvers or Monte-Carlo
simulations to analyze the transport equation is a complex procedure
that is inconvenient from the point of view of X-ray data analysis.
Furthermore, the results obtained for the spectra often depend on
boundary conditions in the energy coordinate that are not known with
precision a priori. In order to limit the effect of the imprecisely
known boundary conditions on the computed spectrum, one is usually
forced to extend the computational domain well beyond the region of
interest. The associated decrease in computational efficiency reduces
the utility of simulations based on numerical integration of the
transport equation. Moreover, even when accurate numerical solutions can
be obtained, it is often difficult to extract simple analytical
estimates from them due to the complexity of the simulations. 

Our increasing capability to make observations with high temporal and
spectral resolution presents us with an interesting theoretical
challenge, which is to obtain a closed-form expression for the X-ray
spectrum resulting from the thermal Comptonization of initially
monoenergetic photons in an ionized plasma. The availability of such a
solution would be of great importance for our understanding of spectral
formation during rapid X-ray transients, and particularly when the
Compton broadening of narrow features such as iron lines is of interest.
As a first step, in this paper we analyze the transport equation for
thermal Comptonization and derive the time-dependent Green's function
for the case of a homogeneous plasma with steady properties. Our goal is
to obtain the ``complete'' Green's function, including the effect of
electron recoil as well as the diffusion of photons in the energy space.
The formalism also incorporates a treatment of spatial transport inside
the cloud. The resulting solution can be used to obtain a comprehensive
understanding of the spectral evolution due to thermal Comptonization in
a plasma with any shape and with any temporal and spatial distribution
of embedded photon sources. We believe that the insights gained using
this new solution will be valuable for the interpretation of current and
future X-ray data from active galaxies and low-mass X-ray binaries.

The remainder of the paper is organized as follows. In \S~2 we introduce
and discuss the fundamental transport equation governing the propagation
of photons in X-ray emitting plasma clouds. In \S~3 the transport
equation is solved to obtain the formal solution for the time-dependent
Green's function in the energy domain as an integral of a product of two
Whittaker functions. The properties of the Green's function are explored
in \S~4, where we also obtain a particular solution describing the
evolution a bremsstrahlung initial spectrum including a low-energy
cutoff. Specific examples of the time-dependent energy spectra emitted
by Comptonizing clouds are computed in \S~5, and the implications of our
results for the interpretation of X-ray data are discussed in \S~6.
Supplemental technical details of the mathematical approach are provided
in Appendix~A.

\section{TIME-DEPENDENT COMPTONIZATION}

In radiation-dominated, fully-ionized plasmas, photon creation and
destruction are unable to establish local thermodynamic equilibrium, and
the photons and electrons interact primarily via Compton scattering.
In this section, we discuss the general equations governing the photon
distribution as a function of time, space, and energy. The evolution
equation for the photon energy distribution is solved in \S~3.

\subsection{Transport equation}

Neglecting absorption, the temporal evolution of the distribution of
photons in a nonrelativistic plasma cloud with electron number density
$n_e$ and electron temperature $T_e$ is governed by the transport
equation (Katz 1976; Payne 1980)
\begeq
{\partial \oc \over \partial t}
= {n_e\sig c\over m_e c^2}\,{1\over\epsilon^2} \,
{\partial \over \partial \epsilon} \left[\epsilon^4 \left(\oc
+ \oc^2 + k T_e \, {\partial \oc \over \partial \epsilon}
\right)\right] + \vec\nabla \cdot\left({c \over 3 \, n_e \sig}
\, \vec\nabla \oc\right) + j \ ,
\label{eq1}
\fineq
where $\oc(\epsilon,\vec r,t)$ is the photon occupation number (a
dimensionless quantity), $\epsilon$ is the photon energy, $t$ is time,
$\vec r$ is the spatial location within the plasma, $\sig$ is the
Thomson cross section, $c$ is the speed of light, $m_e$ is the electron
mass, and $k$ is Boltzmann's constant. The specific intensity, $I_\nu
\propto {\rm ergs \ cm^{-2} \, s^{-1} \, ster^{-1} \, Hz^{-1}}$, is
related to the occupation number by $I_\nu = (2 h \nu^3/c^2)\,\oc$,
where $h$ is Planck's constant and $\nu = \epsilon/h$ is the photon
frequency. The terms on the right-hand side of equation~(\ref{eq1})
represent thermal Comptonization, spatial diffusion, and photon sources,
respectively. We shall assume that $I_\nu$, $\oc$, and the source term
$j$ are all isotropic. In this paper we will focus primarily on the
process of thermal Comptonization occurring in homogeneous plasmas with
steady properties. Effects related to bulk motion and temperature
variations will be discussed in \S~6.

The occupation number $\oc$ can be integrated with respect to $\epsilon$
to obtain the total radiation number density, given by
\begeq
n_r(\vec r,t) = \int_0^\infty {8 \, \pi \over c^3 h^3} \ \epsilon^2
\, \oc(\epsilon, \vec r,t) \, d\epsilon \ .
\label{eq2}
\fineq
The $\oc^2$ term in equation~(\ref{eq1}) describes stimulated scattering
and is rarely important in astrophysical applications. When this term is
neglected, the transport equation is rendered linear, and in the case of
an isothermal plasma, we can obtain the equivalent form
\begeq
{\partial \oc \over \partial t} =
{n_e \sig c \ k T_e \over m_e c^2} \, {1 \over x^2} {\partial \over
\partial x} \left[x^4 \left(\oc + {\partial \oc \over \partial x}
\right)\right] + \vec\nabla \cdot\left({c \over 3 \, n_e \sig} \,
\vec\nabla \oc\right) + j \ ,
\label{eq2a}
\fineq
where we have introduced the dimensionless photon energy
\begeq
x(\epsilon) \equiv {\epsilon \over k T_e} \ .
\label{eq2b}
\fineq
By transforming the variable of integration in equation~(\ref{eq2})
from $\epsilon$ to $x$, we can reexpress the photon number density
as
\begeq
n_r(\vec r,t) = 8 \pi \left(k T_e \over c h\right)^3 \int_0^\infty
\ x^2 \, \oc(x, \vec r,t) \, dx \ .
\label{eq2c}
\fineq
Since equation~(\ref{eq2a}) is linear in $\oc$, it is sufficient to
consider a source term $j=j_*$ that is localized in time, space, and
energy, given by
\begeq
j_* \equiv {1 \over 8 \pi x_0^2} \left(h c \over k T_e \right)^3
\, \delta(t-t_0) \, \delta(\vec r - \vec r_0) \, \delta(x-x_0) \ ,
\label{eq2d}
\fineq
which represents the injection of a single photon with dimensionless
energy $x_0 = \epsilon_0/(kT_e)$ at time $t_0$ and location $\vec r_0$.
The solution to equation~(\ref{eq2a}) corresponding to the soure term $j_*$
is the Green's function, $\greenoc(x,x_0,\vec r,\vec r_0,t,t_0)$, describing
the process of time-dependent thermal Comptonization in an isothermal plasma.

Once the Green's function has been determined by solving equations~(\ref{eq2a})
and (\ref{eq2d}) in combination with a suitable boundary condition imposed at
the surface of the cloud, the response to a general source term $j$ can be
obtained by performing the convolution
\begeq
\oc(x, \vec r,t) = 8 \pi \left(k T_e \over c h\right)^3
\oint_V\int_0^t\int_0^\infty \greenoc(x,x_0,\vec r,\vec r_0,t,t_0) \,
j(x_0,t_0,\vec r_0) \, x_0^2 \, dx_0 \, dt_0 \, d^3 \vec r_0 \ ,
\label{convolve1}
\fineq
where the $\vec r_0$ integration proceeds over the entire volume
$V$ of the cloud, regardless of its shape. Equation~(\ref{convolve1})
implicitly assumes that no radiation is present in the cloud at time
$t=0$, and that no radiation is incident on the cloud from the outside.
As an alternative to solving for $\greenoc$ using equations~(\ref{eq2a})
and (\ref{eq2d}), we point out that $\greenoc$ is also the solution to
the {\it homogeneous} equation
\begeq
{\partial \greenoc \over \partial t} =
{n_e \sig c \ k T_e \over m_e c^2} \, {1 \over x^2} {\partial \over
\partial x} \left[x^4 \left(\greenoc + {\partial \greenoc \over \partial x}
\right)\right] + \vec\nabla \cdot\left({c \over 3 \, n_e \sig} \,
\vec\nabla \greenoc\right) \ ,
\label{transport1}
\fineq
subject to the {\it initial condition}
\begeq
\greenoc(x,x_0,\vec r,\vec r_0,t,t_0) \Big|_{t=t_0} = {1 \over 8 \pi x_0^2}
\left(h c \over k T_e \right)^3 \, \delta(\vec r - \vec r_0)
\, \delta(x-x_0) \ .
\label{eq4}
\fineq
The homogeneity of equation~(\ref{transport1}) makes it a more convenient
starting point than equation~(\ref{eq2a}) for the determination of the
Green's function $\greenoc$. In a homogeneous plasma with steady properties,
both $T_e$ and $n_e$ are constants, and it is convenient to rewrite
equation~(\ref{transport1}) as
\begeq
{\partial \greenoc \over \partial y} =
{1 \over x^2} {\partial \over \partial x} \left[x^4 \left(\greenoc
+ {\partial \greenoc \over \partial x}\right)\right] + \vec\nabla
\cdot\left({c \over 3 \, \alpha \, n_e \sig} \, \vec\nabla
\greenoc\right) \ ,
\label{transport2}
\fineq
where we have introduced the dimensionless time
\begeq
y(t) \equiv \, \alpha \, (t-t_0) \ , \ \ \ \ \ 
\alpha \equiv n_e \sig c \, {k T_e \over m_e c^2} \ .
\label{eq11}
\fineq
This is the familiar Compton $y$-parameter, which generally must
exceed unity in order for significant modification of the spectrum
to occur (Rybicki \& Lightman 1979). The constant $\alpha$ is the
``Comptonization rate,'' and $y=0$ at the initial time $t=t_0$.
In the case of a homogeneous plasma with constant density and
temperature, the temporal evolution of the photon distribution
depends only on the elapsed time since injection, $t-t_0$, and
therefore we can write $\greenoc(x,x_0,\vec r,\vec r_0,t,t_0)
=\greenoc(x,x_0,\vec r,\vec r_0,y)$ without loss of generality.

By operating on equation~(\ref{transport2}) with $8 \pi (kT_e/hc)^3
\int_0^\infty x^2\,dx$, we can show that the radiation number
density associated with the Green's function,
\begeq
\greendens(\vec r,\vec r_0,y) \equiv 8 \pi \left(k T_e \over c h\right)^3
\int_0^\infty \ x^2 \, \greenoc(x,x_0,\vec r,\vec r_0,y)
\, dx \ ,
\label{eq11b}
\fineq
satisfies the homogeneous spatial diffusion equation
\begeq
{\partial \greendens \over \partial y} = \vec\nabla\cdot\left(
{c \over 3 \, \alpha \, n_e \sig} \, \vec\nabla \greendens \right) \ .
\label{eq5}
\fineq
Note that the Comptonization term in equation~(\ref{transport2})
vanishes upon integration, which is a manifestation of the fact that
Compton scattering does not create or destroy photons. The initial
condition for $\greendens$ can be obtained by combining
equations~(\ref{eq4}) and (\ref{eq11b}), which yields
\begeq
\greendens(\vec r,\vec r_0,y) \Big|_{y=0} = \delta(\vec r - \vec r_0)
\ .
\label{eq6}
\fineq
Note that equations~(\ref{eq5}) and (\ref{eq6}) must be supplemented by
a suitable flux boundary condition imposed at the surface of the cloud
in order to account for the escape of radiation from the plasma.

\subsection{Separability}

We shall focus on the evolution of the occupation number Green's
function $\greenoc$ subject to the initial condition given by
equation~(\ref{eq4}). The transport equation~(\ref{transport2})
is separable in time, space, and energy when $n_e$ and $T_e$ are
constants, as assumed here (Payne 1980). We employ separability
by writing the occupation number $\oc$ as the product
\begeq
\greenoc(x,x_0,\vec r,\vec r_0,y) = {1 \over 8 \pi} \left(
h c \over k T_e
\right)^3 \ \greendens(\vec r,\vec r_0,y) \ \green(x,x_0,y) \ ,
\label{eq9}
\fineq
where $\green(x,x_0,y)$ is the Green's function for the photon energy
distribution. Note that the energy and spatial dependences have been
separated since the former is contained in $\green$ and the latter is
contained in $\eta$. Substituting for $\greenoc$ in the transport
equation~(\ref{transport2}) using equation~(\ref{eq9}), and utilizing
the fact that $\eta$ satisfies equation~(\ref{eq5}), we can show that
the energy distribution $\green$ satisfies the Kompaneets (1957)
Fokker-Planck equation,
\begeq
{\partial \green \over \partial y} = {1 \over x^2} {\partial \over
\partial x} \left[x^4 \left(\green + {\partial \green \over \partial x}
\right)\right] \ ,
\label{eq16}
\fineq
along with the initial condition
\begeq
\green(x,x_0,y) \Big|_{y=0} = x_0^{-2} \, \delta(x-x_0)
\label{eq17} \ .
\fineq
The terms proportional to $\green$ and $\partial \green / \partial x$
inside the parentheses on the right-hand side of equation~(\ref{eq16})
express the effects of electron recoil and stochastic (second-order
Fermi) photon energization, respectively. Due to our neglect of
stimulated scattering, the asymptotic solution to equation~(\ref{eq16})
obtained as $y \to \infty$ is the Wien spectrum $\green \propto e^{-x}$,
rather than a Bose-Einstein distribution. The function $\green(x,x_0,y)$
represents the solution to the Kompaneets equation in an infinite,
homogeneous medium. Spatial effects associated with the geometry and the
finite size of the scattering cloud are introduced via the density
distribution $\eta(\vec r,\vec r_0,y)$. Note that $\greenoc$ satifies
the initial condition given by equation~(\ref{eq4}) as required.

Once the spatial distribution has been obtained by solving
equation~(\ref{eq5}) for $\eta(\vec r,\vec r_0,y)$, we must next
determine the energy distribution $\green(x,x_0,y)$ in order to
construct the complete solution for $\greenoc$ using
equation~(\ref{eq9}). Although Payne (1980) and Rybicki \& Lightman
(1979) state that equation~(\ref{eq16}) must be solved numerically in
general, we shall demonstrate below that an analytical solution for the
Green's function can in fact be obtained in the form of a real integral.
By operating on equation~(\ref{eq16}) with $\int_0^\infty x^2 \, dx$, we
can establish that $\green$ has the convenient normalization
\begeq
\int_0^\infty x^2 \, \green(x,x_0,y) \, dx
= {\rm constant} = 1 \ ,
\label{eq19}
\fineq
where the final result follows from the initial condition
(eq.~[\ref{eq17}]). Note that this normalization is maintained for all
values of $y$, which reflects the fact that Compton scattering conserves
photons. We shall seek to solve equation~(\ref{eq16}) using Laplace
transformation in \S~3. Once the solution for the Green's function
$\green(x,x_0,y)$ is known, the particular solution for the distribution
function $f(x,y)$ corresponding to an arbitrary initial spectrum
$f_0(x)$ can be found using the integral formula
\begeq
f(x,y) = \int_0^\infty x_0^2 \, f_0(x_0) \, \green(x,x_0,y)
\, dx_0 \ .
\label{convolve}
\fineq
By operating on equation~(\ref{convolve}) with $\int_0^\infty x^2
\, dx$, we can establish that
\begeq
I_2(y) \equiv \int_0^\infty x^2 \int_0^\infty x_0^2 \, f_0(x_0)
\, \green(x,x_0,y) \, dx_0 \, dx
= \int_0^\infty x_0^2 \, f_0(x_0) \, dx_0 = {\rm constant} \ ,
\label{convolve2}
\fineq
where the final result is obtained by reversing the order of integration
and applying equation~(\ref{eq19}). The constancy of $I_2$ follows from
the fact that Comptonization does not create or destroy photons, and we
therefore refer to $I_2$ as the ``number moment'' of the particular
solution $f(x,y)$. Equation~(\ref{convolve2}) will provide a useful
check when we consider the properties of the particular solutions
obtained in \S~4.

\section{SOLUTION FOR THE GREEN'S FUNCTION}

Our analytical approach to the determination of the Green's function
energy distribution $\green(x,x_0,y)$ will be based on Laplace transformation.
For the sake of brevity, only the primary steps are discussed here. Readers
interested in the technical details are referred to Appendix~A.

\subsection{Laplace transformation}

Laplace transformation of the Kompaneets partial differential
equation~(\ref{eq16}) with respect to $y$ yields the ordinary
differential equation
\begeq
s \, L - x_0^{-2} \, \delta(x-x_0) = {1 \over x^2} {d \over dx}
\left[x^4 \left(L + {dL \over dx}\right)\right] \ ,
\label{eq20}
\fineq
where
\begeq
L(x,x_0,s) \equiv \int_0^\infty e^{-s y} \, \green(x,x_0,y) \, dy
\label{eq21}
\fineq
denotes the Laplace transform of $\green(x,x_0,y)$, and we have also used
equation~(\ref{eq17}). We show in Appendix~A.1 that the solution for the
transform $L(x,x_0,s)$ is given by
\begeq
L(x,x_0,s) = {\Gamma(\mu-3/2) \over \Gamma(1+2\mu)} \,
x_0^{-2} \, x^{-2} \, e^{(x_0-x)/2} \,
\cases{
W_{2, \, \mu}(x_0) \, M_{2, \, \mu}(x) \ , & $x \le x_0$ \ , \cr
\phantom{stuff} \cr
M_{2, \, \mu}(x_0) \, W_{2, \, \mu}(x) \ , & $x \ge x_0$ \ , \cr
}
\label{eq22}
\fineq
where the quantity $\mu$ is a function of the transform variable $s$,
defined by
\begeq
\mu(s) \equiv \left(s + {9 \over 4}\right)^{1/2} \ ,
\label{eq25}
\fineq
and $M_{2, \, \mu}(x)$ and $W_{2, \, \mu}(x)$ denote the Whittaker functions
(Abramowitz \& Stegun 1970). Equation~(\ref{eq22}) can be rewritten in
the equivalent form
\begeq
L(x,x_0,s) = {\Gamma(\mu-3/2) \over \Gamma(1+2\mu)} \,
x_0^{-2} \, x^{-2} \, e^{(x_0-x)/2} \,
M_{2, \, \mu}(\xmin) \ W_{2, \, \mu}(\xmax) \ ,
\label{eq23}
\fineq
where
\begeq
\xmin \equiv \min(x,x_0) \ , \ \ \ \ \ \ 
\xmax \equiv \max(x,x_0) \ .
\label{eq24}
\fineq

The function $W_{2, \, \mu}(x)$ is defined in terms of $M_{2, \, \mu}(x)$
by equation~(13.1.34) of Abramowitz \& Stegun (1970), which gives
\begeq
W_{2, \, \mu}(x) \equiv {\Gamma(-2\mu) \over \Gamma(-\mu-3/2)}
\, M_{2, \, \mu}(x) + {\Gamma(2\mu) \over \Gamma(\mu-3/2)}
\, M_{2, \, -\mu}(x) \ ,
\label{whit1}
\fineq
and the function $M_{2, \, \mu}(x)$ can be evaluated using the series
expansion
\begeq
M_{2, \, \mu}(x) = e^{-x/2} \, x^{\mu+1/2} \, \left[1 + {\mu-3/2 \over
1+2\mu} \, x + {(\mu-3/2)(\mu-1/2) \over (1+2\mu)(2+2\mu)}
\, {x^2 \over 2} + \cdots \right] \ .
\label{whit2}
\fineq
The solution for $L(x,x_0,s)$ in equation~(\ref{eq23}) has been obtained
by utilizing continuity and derivative jump conditions based on the
differential equation~(\ref{eq20}), along with various identities
satisfied by the Whittaker functions. Further details of the derivation
are provided in Appendix~A.1.

\subsection{Integral expression for the Green's function}

To obtain the solution for the Green's function $\green(x,x_0,y)$, we must
perform the inverse Laplace transformation of $L(x,x_0,s)$ using the complex
Mellin inversion integral (Butkov 1968),
\begeq
\green(x,x_0,y) = {1 \over 2 \pi i} \int_{\gamma-i\infty}^{\gamma+i\infty}
e^{sy} \, L(x,x_0,s) \, ds \ ,
\label{eq26}
\fineq
where the constant $\gamma$ is selected so that the line ${\rm Re} \, s
= \gamma$ lies to the right of any singularities in the integrand. In
this problem, singularities occur where the quantity $\mu-3/2$ is zero
or a negative integer, resulting in the divergence of $\Gamma(\mu-3/2)$.
The definition of $\mu$ (eq.~[\ref{eq25}]) therefore implies that
simple poles are located at $s=0$ and $s=-2$. The corresponding values
for $\mu$ are $\mu=3/2$ and $\mu=1/2$, respectively. From the locations
of the singularities, it follows that we must require $\gamma > 0$ for
convergence of the integral in equation~(\ref{eq26}).

The simple poles at $s=0$ and $s=-2$ both lie to the right of the branch
point for the square root function, which is located at $s = -9/4$. Hence
they are contained within the closed integration contour $C$ indicated in
Figure~1. We can therefore use the residue theorem to write
\begeq
\oint_C e^{sy} \, L(x,x_0,s) \, ds
= 2 \pi i \, \sum_{n=1}^2 \, {\rm Res}(s_n) \ ,
\label{eq31}
\fineq
where the left-hand side denotes the integral around the contour $C$ and
${\rm Res}(s_n)$ is the residue associated with the simple pole located
at $s = s_n$, with $s_1=0$ and $s_2=-2$. Note that the integration
contour must avoid the branch cut of the square root function,
extending from $s = -9/4$ to $s = - \infty$. Asymptotic analysis
indicates that the contributions to the integral along the large
arcs $MN$ and $QR$ vanish in the limit $r_1 \to \infty$ (see Fig.~1).
Likewise, the integration along the small arc $OP$ vanishes in the
limit $r_2 \to 0$. Our solution for the energy distribution $\green$
therefore reduces to
\begeq
\green(x,x_0,y) =
- {1 \over 2\pi i} \int_N^O e^{sy} \, L(x,x_0,s) \, ds
- {1 \over 2\pi i} \int_P^Q e^{sy} \, L(x,x_0,s) \, ds
+ \sum_{n=1}^2 \, {\rm Res}(s_n) \ ,
\label{eq32}
\fineq
in the limit $r_1 \to \infty$, $r_2 \to 0$. Equation~(\ref{eq32}) expresses
the Laplace inversion in terms of two residues and two integrals, one above
the branch cut and one below it.

\begin{figure}
\hspace{50mm}
\includegraphics[width=85mm]{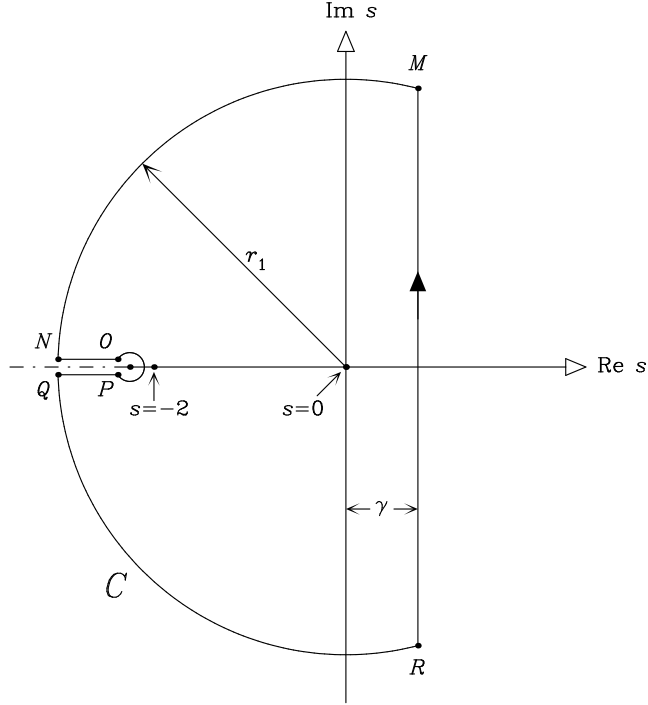}
\caption{Closed integration contour $C$ for the Laplace inversion
integral on the left-hand side of equation~(\ref{eq31}). The branch
point of the complex square root function is located at $s=-9/4$, in the
center of the small arc $OP$, which has radius $r_2$. Simple poles are
located at $s=0$ and $s=-2$; these singularities are contained within
the contour $C$ provided $\gamma > 0$. The branch cut (dot-dashed line)
extends from the branch point to $s = - \infty$. See the discussion in
the text.}
\end{figure}

The residues appearing in equation~(\ref{eq32}) are evaluated in
Appendix~A.2, where we find that (see eqs.~[\ref{ac6}])
\begeq
\sum_{n=1}^2 \, {\rm Res}(s_n) = {e^{-x} \over 2}
+ {e^{-x-2y} \over 2} \ {(2 - x) \, (2 - x_0) \over
\ x_0 \, x} \ .
\label{eq33}
\fineq
The procedure for treating the integrations above and below the branch
cut is discussed in Appendix~A.3. This involves the utilization of
symmetry relations and other identities satisfied by the Whittaker
functions. The final result given by equation~(\ref{ac16}) is
\begin{eqnarray}
\green(x,x_0,y) =
{32 \over \pi} \ e^{-9y/4} x_0^{-2} x^{-2} e^{(x_0-x)/2}
\int_0^\infty e^{-u^2 y} \, {u \, \sinh(\pi u) \over
(1 + 4 u^2)(9 + 4 u^2)} \phantom{SPAAAAAACE} \nonumber \\
\times \ W_{2, \, i u}(x_0) \,
W_{2, \, i u}(x) \, du
\ + \ {e^{-x} \over 2}
\ + \ {e^{-x-2y} \over 2} \ {(2 - x) \, (2 - x_0) \over
\ x_0 \, x} \ .
\label{eq34}
\end{eqnarray}
This solution for the Green's function is one of the main results of the
paper. It represents the fundamental ``kernel'' for the process of
thermal Comptonization in an infinite homogeneous medium, from which
particular solutions for any distribution of photon sources can be
developed by quadrature. Spatial effects in a finite medium can be
incorprated using the separation solution given by equation~(\ref{eq9}).
In Appendix~A.4, we provide the series expansions needed to evaluate the
Whittaker functions in equation~(\ref{eq34}). Furthermore, a
self-contained FORTRAN code that evaluates the Green's function by
performing the integration in equation~(\ref{eq34}) is available from
the author upon request. In the remainder of the paper, we will analyze
the properties of the Green's function, and demonstrate that all of the
previously known steady-state and time-dependent solutions can be
reproduced using it. Hence equation~(\ref{eq34}) represents a
significant generalization of the earlier analytical results in the
theory of time-dependent thermal Comptonization.

\subsection{Approach to Wien equilibrium}

In the time-dependent, isothermal Comptonization problem treated here,
the Green's function must approach the Wien equilibrium spectrum, i.e.,
\begeq
\lim_{y \to \infty} \green(x,x_0,y)
= \wien(x) \equiv {1 \over 2} \, e^{-x} \ ,
\label{wien1}
\fineq
where the factor of $1/2$ on the right-hand side is required in order to
ensure that $\wien$ satisfies the normalization condition
\begeq
\int_0^\infty x^2 \, \wien(x) \, dx = 1 \ ,
\label{wien2}
\fineq
in compliance with equation~(\ref{eq19}). Our solution for
$\green(x,x_0,y)$ expressed by equation~(\ref{eq34}) clearly satisfies
equation~(\ref{wien1}), and therefore the Green's function exhibits the
correct asymptotic behavior for large values of $y$. In Figure~2 we plot
$\green$ as a function of the dimensionless photon energy $x$ and the
dimensionless time $y$ for four values of the initial photon energy,
$x_0=0.1$, $x_0=1$, $x_0=5$, and $x_0=10$. Also included for comparison
is the Wien equilibrium spectrum, $\wien(x)=(1/2) \, e^{-x}$. Note that
in each case, the Green's function approaches the Wien shape for large
values of $y$ as predicted. The spectrum depicted in Figure~2 describes
the Comptonization of an intrinsically narrow line feature. Applications
to the broad Fe K$\alpha$ lines observed in AGNs are discussed in \S~6.

\begin{figure}
\hspace{15mm}
\includegraphics[width=125mm]{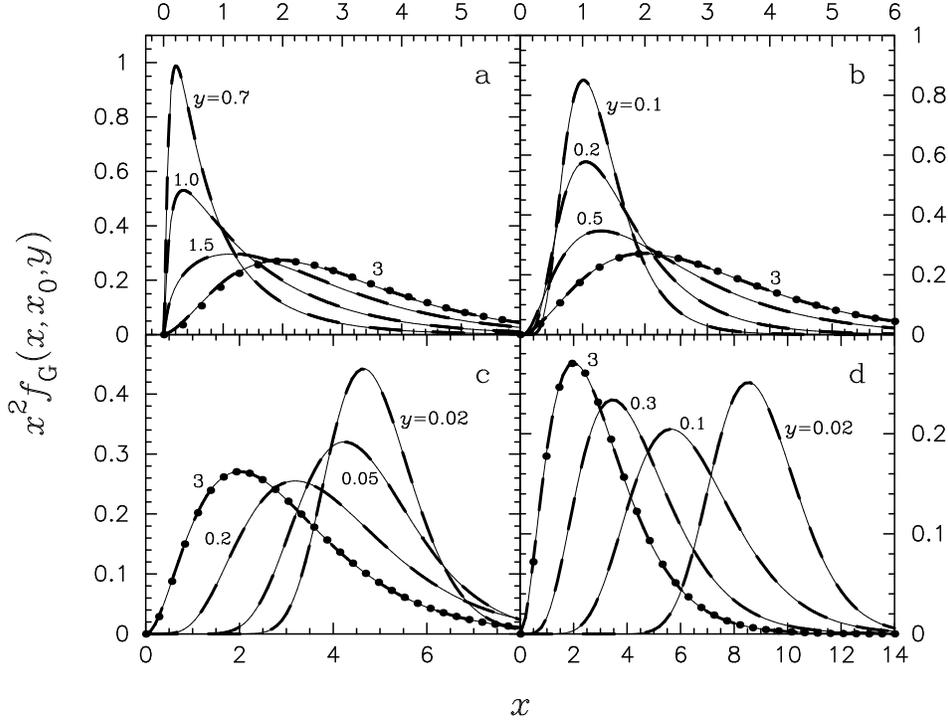}
\caption{Analytical solution for the Green's function $x^2 \green(x,x_0,y)$
(eq.~[\ref{eq34}]; solid lines) plotted as a function of the dimensionless
photon energy $x$ for the indicated values of the dimensionless time $y$.
The area under each curve is equal to unity, and the initial photon energy
is given by (a) $x_0 = 0.1$; (b) $x_0 = 1$; (c) $x_0 = 5$; and (d) $x_0 = 10$.
Included for comparison are the numerical solutions (dashed lines) obtained
by integrating numerically the Kompaneets equation using a Gaussian initial
condition, as explained in the text. The two sets of curves are essentially
identical. As $y$ increases, $x^2 \green(x,x_0,y)$ approaches the Wien spectrum
$(1/2) \, x^2 e^{-x}$ (filled circles) in each case as expected.}
\end{figure}

\subsection{Comparison with numerical simulations}

The acid test for our Green's function solution is provided by comparing
it with the photon energy distribution obtained by solving numerically
the Kompaneets partial differential equation~(\ref{eq16}). In addition
to specifying the initial condition for the photon distribution, the
numerical solution procedure also requires the imposition of boundary
conditions for $\green$ at large and small values of $x$, for all values
of $y$. Obviously, this information is not available a priori, and
therefore we are forced to push the boundaries of the computational
domain as far as possible away from the region of interest, so as to
avoid contaminating the solution with imprecise boundary data. This
requirement markedly reduces the efficiency of the numerical approach,
since much of the computer time is spent on calculations outside the
region of interest. We have performed numerical simulations using the
IMSL partial differential equation solving subroutine DMOLCH. The
initial condition is specified as a narrow Gaussian feature, with
selected values for the mean energy $\langle x\rangle$ and standard
deviation $\sigma$. This initial condition approximates the
$\delta$-function distribution given by equation~(\ref{eq17}) when
$x_0=\langle x\rangle$ and $\sigma/\langle x \rangle \ll 1$. The
analytical and numerical solutions for $\green(x,x_0,y)$ are compared in
Figure~2 for four different scenarios. In Figure~2a, we set $\langle
x\rangle=0.1$ and $\sigma=0.01$; in Figure~2b we set $\langle
x\rangle=1.0$ and $\sigma=0.01$; in Figure~2c we set $\langle
x\rangle=5.0$ and $\sigma=0.05$; and in Figure~2d we set $\langle
x\rangle=10$ and $\sigma=0.1$. The results are plotted for several
values of $y$. Note that the numerical and analytical solutions are
virtually indistinguishable, confirming that our closed-form expression
(eq.~[\ref{eq34}]) is in fact an exact representation of the Green's
function.

\section{PROPERTIES OF THE GREEN'S FUNCTION}

The solution we have obtained in \S~3.2 for the Green's function
$\green(x,x_0,y)$ given by equation~(\ref{eq34}) has a number of
interesting properties that are further explored in this section.

\subsection{Moments of the Green's function}

We can obtain additional insight into the behavior of the Green's
function by focusing on the variation of the ``power moments,''
$\greenmom_n(y)$, defined by
\begeq
\greenmom_n(y) \equiv \int_0^\infty x^n \, \green(x,x_0,y) \, dx \ .
\label{moment1}
\fineq
Although the results obtained below for $\greenmom_n(y)$ are valid for
general real values of $n$, in our specific applications we shall focus
on integral values, since these correspond to cases of special physical
interest such as the ``number moment'' $\greenmom_2$ introduced in
equation~(\ref{convolve2}), and the ``energy moment'' $\greenmom_3$ discussed
below. In the case of the Green's function, the initial values of the power
moments at $y=0$ are obtained by substituting equation~(\ref{eq17}) into
equation~(\ref{moment1}), which yields
\begeq
\greenmom_n(0) = x_0^{n-2} \ .
\label{moment2}
\fineq
We can determine the subsequent evolution of $\greenmom_n(y)$ by operating
on equation~(\ref{eq34}) with $\int_0^\infty x^n \, dx$. The result obtained
is
\begin{eqnarray}
\greenmom_n(y) = {32 \over \pi} \ e^{-9y/4} x_0^{-2} e^{x_0/2}
\int_0^\infty e^{-u^2 y} \, {u \, \sinh(\pi u) \over
(1 + 4 u^2)(9 + 4 u^2)} \, W_{2, \, i u}(x_0)
\phantom{SPAAAAAACE} \nonumber \\
\times \ \int_0^\infty e^{-x/2}
x^{n-2} \, W_{2, \, i u}(x) \, dx \ du
+ {\Gamma(n+1) \over 2} + (2-n) \, \Gamma(n) \, e^{-2y}
\left({1 \over x_0} - {1 \over 2} \right) \ ,
\label{moment3}
\end{eqnarray}
where we have interchanged the order of integration over $u$ and
$x$. The integration with respect to $x$ converges provided $n > 1/2$,
in which case we can use equation~(7.621.11) from Gradshteyn \& Ryzhik
(1980) to find that
\begin{eqnarray}
\greenmom_n(y) = {32 \, e^{-9y/4} \, e^{x_0/2} \over
\pi \, x_0^2 \, \Gamma(n-2)} \,
\int_0^\infty e^{-u^2 y} \, {u \, \sinh(\pi u)
\over (1 + 4 u^2)(9 + 4 u^2)} \, W_{2, \, i u}(x_0)
\, \Gamma(n-1/2+iu)
\phantom{SPAACE} \nonumber \\
\times \ \Gamma(n-1/2-iu) \, du
\ + \ {\Gamma(n+1) \over 2} + (2-n) \, \Gamma(n) \, e^{-2y}
\left({1 \over x_0} - {1 \over 2} \right)
\ . \phantom{SP}
\label{moment4}
\end{eqnarray}
We shall examine the behavior of this general expression for several
cases of special interest below.
 
\subsection{Variation of the number and energy moments}

We have established via integration of the Kompaneets equation~(\ref{eq16})
that $\green(x,x_0,y)$ must satisfy the normalization condition (cf.
eq.~[\ref{eq19}])
\begeq
\greenmom_2(y) =
\int_0^\infty x^2 \, \green(x,x_0,y) \, dx = 1
\label{moment5}
\fineq
for all values of $y$, in agreement with the initial condition for
$\greenmom_2(y)$ expressed by equation~(\ref{moment2}). As pointed out
in \S~2.2, this behavior is a consequence of the fact that Compton
scattering conserves the photon number density. We can use
equation~(\ref{moment4}) to confirm explicitly that this condition is
actually satisfied by our solution for the Green's function. When $n=2$,
the integral in equation~(\ref{moment4}) vanishes due to the factor
$\Gamma(n-2)$ in the denominator, and we obtain
\begeq
\greenmom_2(y) = 1 \ ,
\label{moment6}
\fineq
in satisfaction of equation~(\ref{moment5}). We have therefore established
that our Green's function solution conserves photons and is properly
normalized. Next we focus on the variation of the ``energy moment,''
\begeq
\greenmom_3(y) \equiv \int_0^\infty x^3 \, \green(x,x_0,y) \, dx \ ,
\label{moment7}
\fineq
which is related to the radiation energy density, $U_r$, via
\begeq
U_r(\vec r,t) = \int_0^\infty {8 \, \pi \over c^3 h^3} \ \epsilon^3
\, \oc(\epsilon, \vec r,t) \, d\epsilon
= 8 \pi {(k T_e)^4 \over (c h)^3} \, \greenmom_3(y) \ .
\label{moment7b}
\fineq
Note that $\greenmom_3(y)$ is also equal to the mean photon energy,
$\bar x(y)$, since
\begeq
\bar x(y) \equiv {\int_0^\infty x^3 \, \green(x,x_0,y) \, dx
\over \int_0^\infty x^2 \, \green(x,x_0,y) \, dx} = \greenmom_3(y) \ ,
\label{moment8}
\fineq
where the final result follows from equation~(\ref{moment6}). The
variation of the mean photon energy was also considered by Kompaneets
(1957), and we shall compare our result for $\bar x(y)$ with his.
Setting $n=3$ in equation~(\ref{moment4}) and using the identity
\begeq
\Gamma(5/2 + iu) \, \Gamma(5/2 - iu)
= {\pi \, (1+4u^2)(9+4u^2) \over 16 \, \cosh(\pi u)} \ ,
\label{moment9}
\fineq
we obtain for the variation of the mean photon energy
\begeq
\bar x(y) = \greenmom_3(y)
= 2 \, x_0^{-2} \, e^{-9y/4} \, e^{x_0/2} \,
\int_0^\infty e^{-u^2 y} \, W_{2, \, i u}(x_0) \ u \,
\tanh(\pi u) \, du
+ 3 - 2 e^{-2y} \, \left({1 \over x_0} - {1 \over 2}\right) \ .
\label{moment10}
\fineq
This result agrees with equation~(48) from Kompaneets (1957),
aside from the fact that his expression for $\bar x(y)$ contains
an erroneous factor of $e^{-x_0/2}$ in place of the correct factor
$e^{x_0/2}$ appearing in equation~(\ref{moment10}). In Figure~3 we
plot the variation of $\bar x(y)$ as a function of $y$ for several values
of the initial photon energy $x_0$. Note that in all cases, we find
that $\bar x(0)=x_0$ initially, as required. For large values of $y$,
we find that $\bar x \to 3$, as expected based on the fact that the
photon distribution should approach the Wien spectrum $\wien(x)
\propto e^{-x}$. However, the evolution of $\bar x$ is not necessarily
monotonic. For example, when $x_0 = 3$, the mean energy $\bar x$
increases until $y \sim 0.25$, and then it decreases, asymptotically
approaching the Wien value of 3 as $y \to \infty$. This behavior can be
understood by examining the associated evolution of the inverse-Compton
temperature, which we investigate in \S~4.3.

\begin{figure}
\hspace{30mm}
\includegraphics[width=100mm]{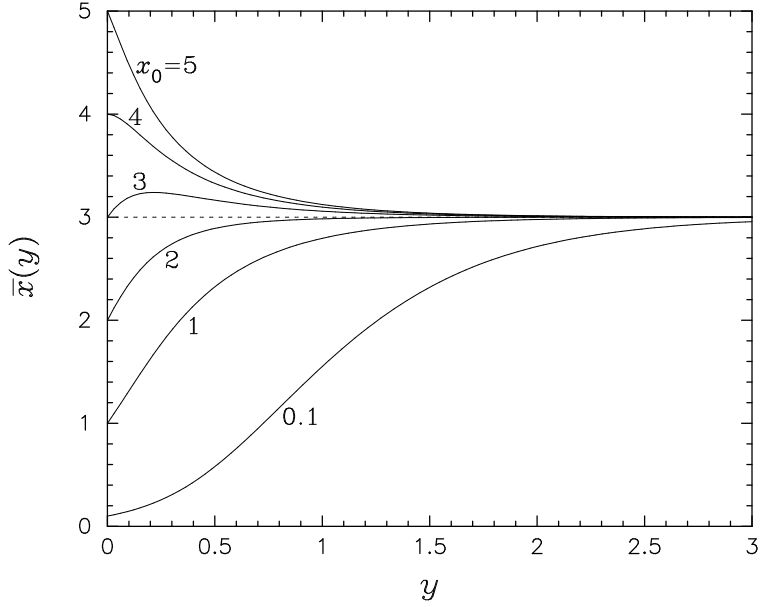}
\caption{Mean photon energy $\bar x$ for the case of a monoenergetic initial
spectrum (eq.~[\ref{moment10}]), plotted as a function of the dimensionless
time $y$ for the indicated values of the initial photon energy $x_0$. Note
that in each case, $\bar x \to 3$ as $y \to \infty$ because the spectrum
always equilibrates to the Wien form. The photons experience continual
heating, on average, when $x_0 \le 2$, and continual cooling when $x_0
\ge 4$. In the intermediate case with $2 < x_0 < 4$, a period of initial
heating is followed by cooling. This behavior is a manifestation of the
underlying variation of the spectral shape, as explained in the text.}
\end{figure}

\subsection{Variation of the inverse-Compton temperature}

By operating on the Kompaneets equation~(\ref{eq16}) with $\int_0^\infty
x^n \, dx$ and integrating by parts twice, we can show that the power
moments $\greenmom_n(y)$ satisfy the ``differential recurrence relation''
\begeq
{d \greenmom_n \over dy} = (n-2) \left[(n+1) \, \greenmom_n(y)
- \greenmom_{n+1}(y)\right]
\ .
\label{IC0}
\fineq
In particular, setting $n=3$ yields the conservation equation for the
energy moment $\greenmom_3(y) = \bar x(y)$,
\begeq
{d \greenmom_3 \over dy} = 4 \, \greenmom_3(y) \left[1
- {\Tic(y) \over T_e}\right]
\ ,
\label{IC1}
\fineq
where $\Tic(y)$ denotes the inverse-Compton temperature of the radiation
field, defined by
\begeq
{\Tic(y) \over T_e} \equiv {1 \over 4} \, {\greenmom_4(y) \over
\greenmom_3(y)} \ ,
\label{IC2}
\fineq
and
\begeq
\greenmom_4(y) \equiv \int_0^\infty x^4 \, \green(x,x_0,y) \, dx \ .
\label{IC3}
\fineq
Equation~(\ref{IC1}) describes the conservation of total energy for a
group of photons experiencing thermal Comptonization in a plasma with a
Maxwellian electron distribution. When $\Tic < T_e$, the photons gain
energy ($\bar x$ increases), and $\bar x$ decreases when $\Tic > T_e$.
The inverse-Compton temperature varies in response to changes in the
shape of the spectrum, and it does not necessarily track the behavior of
$\bar x$. No net energy is exchanged between the radiation and the
matter when $\Tic = T_e$, although subsequent evolution of the spectral
shape can cause $\Tic$ to move away from $T_e$. In the limit $y \to
\infty$, we find that $\Tic \to T_e$ regardless of the initial energy of
the photons because the spectrum asymptotically approaches the Wien form
$\wien \propto e^{-x}$.

In the case of a monoenergetic initial spectrum, we can use
equations~(\ref{moment2}) and (\ref{IC2}) to conclude that
the initial temperature ratio is given by
\begeq
{\Tic(0) \over T_e} = {x_0 \over 4} \ .
\label{IC3b}
\fineq
Hence the photons will initially gain energy if $x_0 < 4$, and otherwise
they will lose energy. The subsequent time evolution of $\Tic(y)$ can be
determined by computing the variation of $\greenmom_4(y)$, which is
accomplished by setting $n=3$ in equation~(\ref{IC0}) and substituting
for $\greenmom_3(y)$ using equation~(\ref{moment10}). The result obtained
is
\begeq
\greenmom_4(y) = {e^{-9y/4} \, e^{x_0/2} \over 2 \, x_0^2}
\, \int_0^\infty e^{-u^2 y} \, W_{2, \, i u}(x_0) \ u \,
\tanh(\pi u) \, (25 + 4u^2) \, du
+ 12 - 12 e^{-2y} \, \left({1 \over x_0} - {1 \over 2}\right) \ .
\label{IC5}
\fineq
It is interesting to note that the same result can also be obtained by
setting $n=4$ in equation~(\ref{moment4}), which provides a useful check
on the self-consistency of our formalism. Taken together,
equations~(\ref{moment10}), (\ref{IC2}), and (\ref{IC5}) provide a new,
closed-form solution describing the variation of the inverse-Compton
temperature for an initially monoenergetic photon distribution. This
result has not appeared previously in the literature, and it can be used
to obtain additional physical insight into the energetics of thermal
Comptonization, as discussed below. Our results for $\greenmom_3$ and
$\greenmom_4$ confirm that $\Tic/T_e \to 1$ as $y \to \infty$, since
$\greenmom_4 \to 12$ and $\greenmom_3 \to 3$. This behavior is
consistent with our expectation that the spectrum should equilibrate to
the Wien form.

In Figure~4 we plot the results obtained for $\Tic(y)$ by combining
equations~(\ref{moment10}), (\ref{IC2}), and (\ref{IC5}) for several
different values of the initial photon energy $x_0$. There are a number
of interesting features in the plots. First, note that for the cases with
$x_0 < 4$, the photons are initially cooler than the electrons (see
eq.~[\ref{IC3b}]), and this explains the initial increase in the mean
energy $\bar x$ displayed in Figure~3. Conversely, when $x_0 > 4$, energy
is initially transferred from the photons to the electrons, and therefore
$\bar x$ decreases. Note, however, that the variations of $\bar x(y)$ and
$\Tic(y)$ are not necessarily monotonic. For example, the case with
$x_0 = 3$ displays initial heating, and $\bar x$ reaches a peak at
$y \sim 0.25$. Beyond this point, the inverse-Compton temperature
exceeds $T_e$, and the photons begin to lose energy. This is
a consequence of the fact that $\bar x$ must eventually decrease to
the Wien value of 3 as $y \to \infty$, and this decrease can only
happen if $\Tic > T_e$. Ultimately, $\Tic$ approaches $T_e$ and
equilibrium is achieved. Hence the behaviors of $\Tic$ and $\bar x$
are fundamentally driven by the underlying evolution of the radiation
spectrum.

For initial photon energies in the range $4 < x_0 < 6$,
equation~(\ref{IC3b}) indicates that $\Tic$ exceeds $T_e$ initially, and
therefore the photons must lose energy to the electrons. However,
despite the fact that $\bar x$ is initially decreasing, we observe that
$\Tic$ actually proceeds to {\it increase}. This apparently paradoxical
behavior stems from the variation of the spectral shape. We can obtain
some useful insight into this phenomenon by computing the initial value
of the temperature derivative, $d\Tic/dy$, in the case of a
monoenergetic initial spectrum. By differentiating equation~(\ref{IC2})
with respect to $y$ and using equation~(\ref{IC0}) to evaluate the
derivatives of the moments, we find that for general values of $y$,
\begeq
{d \over dy}{\Tic(y) \over T_e}
= {3 \over 2} {\greenmom_4 \over \greenmom_3}
- {1 \over 2} {\greenmom_5 \over \greenmom_3}
+ {1 \over 4} \left(\greenmom_4 \over \greenmom_3\right)^2
\ .
\label{IC5b}
\fineq
The initial value of the temperature derivative at $y=0$ can be obtained
by using equation~(\ref{moment2}) to substitute for the moments, which
yields
\begeq
{d \over dy}{\Tic(y) \over T_e}\bigg|_{y=0}
= {x_0 \over 4} \, (6 - x_0) \ .
\label{IC6}
\fineq
This result clearly demonstrates that the inverse-Compton temperature
$\Tic$ initially increases if and only if $x_0 < 6$. On the other hand,
according to equations~(\ref{IC1}) and (\ref{IC3b}), the mean photon
energy $\bar x$ initially increases if and only if $x_0 < 4$. Hence
there is an intermediate regime, $4 < x_0 < 6$, within which $\bar x$
initially decreases, while $\Tic$ simultaneously increases. This
illustrates the fact that $\bar x$ and $\Tic$ are not directly
connected, but each is determined by the underlying evolution of the
spectral shape, which is governed by the Kompaneets equation.

\begin{figure}
\hspace{30mm}
\includegraphics[width=100mm]{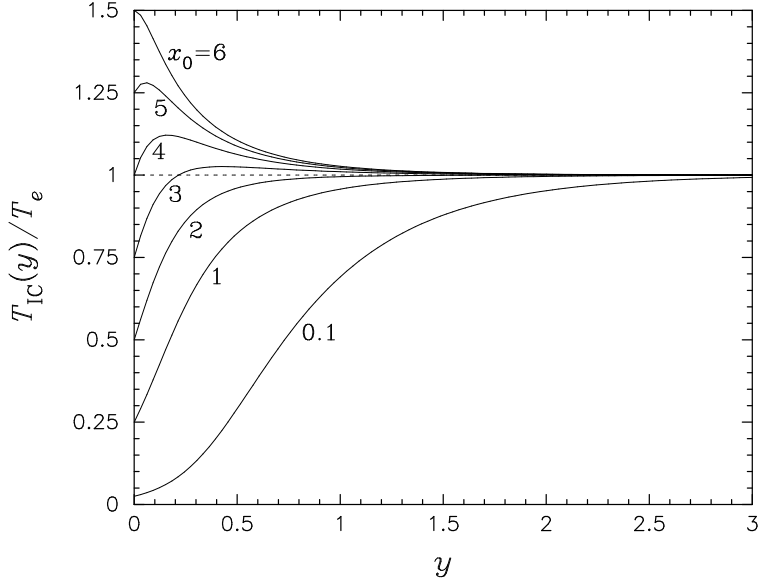}
\caption{Inverse-Compton temperature ratio $\Tic(y)/T_e=(1/4) \,
\greenmom_4(y)/\greenmom_3(y)$ plotted as a function of $y$ for the case
of a monoenergetic initial spectrum with the indicated value of the
initial photon energy $x_0$. The functions $\greenmom_3(y)$ and
$\greenmom_4(y)$ are evaluated using equations~(\ref{moment10}) and
(\ref{IC5}), respectively. The inverse-Compton temperature increases
initially if $x_0 < 6$.}
\end{figure}

\subsection{Particular solution for Wien initial spectrum}

The Green's function solution given by equation~(\ref{eq34}) can be used
to obtain the particular solution $f(x,y)$ corresponding to any desired
initial spectrum $f_0(x)$. As discussed in \S~2.2, the particular solution
is given by the convolution
\begeq
f(x,y) = \int_0^\infty x_0^2 \, f_0(x_0) \, \green(x,x_0,y)
\, dx_0 \ .
\label{wiencon1}
\fineq
The Wien initial spectrum,
\begeq
f_0(x) = e^{-x} \ ,
\label{wiencon2}
\fineq
is a case of special interest, since it represents the equilibrium
solution to the Kompaneets equation~(\ref{eq16}). Substituting
equation~(\ref{wiencon2}) into equation~(\ref{wiencon1}) and
evaluating $\green(x,x_0,y)$ using equation~(\ref{eq34}), we obtain
after interchanging the order of integration over $x_0$ and $u$
\begeq
f(x,y) = e^{-x} \ + \ {32 \over \pi} \ e^{-9y/4} x^{-2} e^{-x/2}
\int_0^\infty e^{-u^2 y} \, {u \, \sinh(\pi u) \over
(1 + 4 u^2)(9 + 4 u^2)} \ W_{2, \, i u}(x)
\int_0^\infty e^{-x_0/2}
W_{2, \, i u}(x_0) \, dx_0 \ du
\ .
\phantom{SPCE}
\label{wiencon3}
\fineq
The integration over $x_0$ in equation~(\ref{wiencon3}) can be performed
using equation~(7.621.11) from Gradshteyn \& Ryzhik (1980), which yields
zero. The particular solution in the case of the Wien initial spectrum
therefore reduces to
\begeq
f(x,y) = e^{-x} \ ,
\label{wiencon5}
\fineq
which confirms that the Wien spectrum remains unaffected by
isothermal Comptonization, as expected.

\subsection{Particular solution for bremsstrahlung initial spectrum}

The reprocessing of optically thin bremsstrahlung emission in clouds of
hot electrons is thought to play a major role in the formation of the X-ray
spectra observed during intense AGN flares (Lightman, Giacconi, \& Tananbaum
1978). Becker \& Begelman (1986) and Chapline \& Stevens (1973) studied
this process by analyzing the evolution of an initial spectrum of the form
\begeq
f_0(x) = x^{-3} \, e^{-x} \ .
\label{bremscon1}
\fineq
The exact particular solution to this problem is given by Becker \& Begelman
as
\begeq
f_{_{\rm BB}}(x,y)
\equiv x^{-3} \, e^{-x} \bigg\{1 + {3 \over 2}\left(1-e^{-2y}\right)
x + {3 \over 2} \left[1 - e^{-2y} \left(2y + 1\right)\right] x^2
+ {3 \over 2} \left[y - 1 + e^{-2y} (y + 1)\right] x^3 \bigg\} \ .
\label{bremscon2}
\fineq
This solution demonstrates the distortion produced by electron
scattering, and the development of a thermal peak in the spectrum.
However, due to the fact that the initial spectrum contains an infinite
number of low-energy photons in this case, it is not possible to achieve
full equilibration to a Wien distribution even in the limit $y \to
\infty$. This unphysical behavior is an artifact of the utilization of
equation~(\ref{bremscon1}) to describe the initial spectrum all the way
down to zero energy. In an actual astrophysical situation, self-absorption
effectively introduces a low-energy cutoff in the initial spectrum,
as discussed below.

During intense AGN flares, soft seed photons produced in a relatively
dense source region via bremsstrahlung are thought to be reprocessed in
a corona containing hot electrons, in which scattering dominates over
photon creation and destruction (Sunyaev \& Titarchuk 1980; Nandra
2001). However, for photon energies below some critical energy $x_*$,
the mean free path for free-free absorption in the source region becomes
smaller than the size of the source. If the source region is
homogeneous, then we can use the usual formulas for thermal
bremsstrahlung to show that (Rybicki \& Lightman 1979)
\begeq
{x^3_* \over 1 - e^{-x_*}} = 6.6 \, {U_r \over a T^4} \ ,
\label{bremscon2b}
\fineq
where $T$ and $U_r$ denote the gas temperature and radiation energy
density in the source region, respectively, and $a T^4$ is the blackbody
energy density. In the physical applications of interest here, $U_r \ll
a T^4$, and therefore we can rewrite equation~(\ref{bremscon2b}) as
\begeq
x_* = 2.6 \, \left(T_{\rm eff} \over T_e\right)^2 \ll 1 \ ,
\label{bremscon2c}
\fineq
where $T_{\rm eff} = (U_r / a)^{1/4}$ is the effective temperature of
the radiation in the source region. Below the critical energy $x_*$, the
initial spectrum given by equation~(\ref{bremscon2}) transitions into a
Planck distribution at temperature $T$ due to self-absorption. Since the
number of photons per unit frequency range contained in the Planck
distribution is proportional to the frequency $\nu$ as $\nu \to 0$, this
effectively introduces a cutoff in the initial spectrum at energy $x =
x_*$. The existence of this low-energy cutoff has a profound effect on
the spectral evolution in the scattering region because the number of
photons contained in the initial spectrum is now {\it finite}.

With the availability of our analytical expression for the Green's
function given by equation~(\ref{eq34}), we are in a position to improve
the situation by treating an optically thin bremsstrahlung initial
spectrum that includes a low-energy cutoff. Specifically, we shall
analyze the evolution of the ``modified'' bremsstrahlung initial
spectrum given by
\begeq
f_0(x) = \cases{
0 \ , & $x < x_*$ \ , \cr
x^{-3} \, e^{-x} \ , & $x \ge x_*$ \ , \cr
}
\label{bremscon3}
\fineq
where $x_* < 1$ denotes the low-energy cutoff, which approximates
the effect of self-absorption. The number moment associated with
this initial spectrum is (see eq.~[\ref{convolve2}])
\begeq
I_2 = \int_0^\infty x^2 \, f_0(x) \, dx = \Gamma(0,x_*) \ ,
\label{bremscon3b}
\fineq
where $\Gamma(a,z)$ denotes the incomplete gamma function (Abramowitz
\& Stegun 1970). Note that the modified bremsstrahlung initial spectrum
contains a finite number of photons. Combining equations~(\ref{eq34}),
(\ref{wiencon1}), and (\ref{bremscon3}), and interchanging the order
of integration over $x_0$ and $u$, we obtain
\begeq
f(x,y) = {32 \over \pi} \ e^{-9y/4} x^{-2} e^{-x/2}
\int_0^\infty e^{-u^2 y} \, {u \, \sinh(\pi u) \over
(1 + 4 u^2)(9 + 4 u^2)} \ W_{2, \, i u}(x)
\int_{x_*}^\infty x_0^{-3} \, e^{-x_0/2}
W_{2, \, i u}(x_0) \, dx_0 \ du
\ + \ S \ ,
\label{bremscon4}
\fineq
where
\begin{eqnarray}
S \equiv {1 \over 2} \, e^{-x} \, \Gamma(0,x_*)
+ e^{-x-2y} \, \left({1 \over x} - {1 \over 2}\right)
\left[2 \, \Gamma(-1,x_*) - \Gamma(0,x_*)\right] \ .
\label{bremscon5}
\end{eqnarray}
The integration over $x_0$ in equation~(\ref{bremscon4}) can be worked
out by employing equation~(13.1.33) from Abramowitz \& Stegun (1970) and
equation~(3.2.12) from Slater (1960) and integrating by parts three times.
After some simplification, we obtain for the particular solution
\begin{eqnarray}
f(x,y) = {32 \over \pi} \ e^{-9y/4} x^{-2} x_*^{-2} e^{-(x+x_*)/2}
\int_0^\infty e^{-u^2 y} \, {u \, \sinh(\pi u) \over
(1 + 4 u^2)(9 + 4 u^2)} \ W_{2, \, i u}(x) \phantom{SPAAAAAACE}
\nonumber \\
\times \left[W_{1, \, i u}(x_*)
- 3 W_{0, \, i u}(x_*) + 6 W_{-1, \, i u}(x_*)
- 6 W_{-2, \, i u}(x_*)\right] \, du
\ + \ S \ .
\label{bremscon6}
\end{eqnarray}
This result describes the time-dependent Comptonization of an optically
thin bremsstrahung initial spectrum with a low-energy cutoff at $x=x_*$
and a finite number of photons. Note that as $y \to \infty$, the integral
vanishes and $S \to \Gamma(0,x_*) \, e^{-x}/2$, and therefore we find
that
\begin{eqnarray}
\lim_{y \to \infty} f(x,y) = {1 \over 2} \, e^{-x} \, \Gamma(0,x_*)
= {I_2 \over 2} \, e^{-x} \ ,
\label{bremscon6b}
\end{eqnarray}
where $I_2 = \Gamma(0,x_*)$ according to equation~(\ref{bremscon3b}).
Hence our solution clearly equilibrates asymptotically to the Wien
spectrum $\wien(x) = (I_2/2) \, e^{-x}$, which satisfies the
normalization requirement $\int_0^\infty x^2 \wien(x) \, dx = I_2$.
Series expansions useful for evaluating the Whittaker functions
appearing in equation~(\ref{bremscon6}) are provided in Appendix~A.4,
and a self-contained FORTRAN code that performs the integration to
determine $f(x,y)$ is available from the author upon request.

The particular solution given by equation~(\ref{bremscon6}) represents
a generalization of the $x_*=0$ result $f_{_{\rm BB}}(x,y)$
(eq.~[\ref{bremscon2}]), and we find that
\begin{eqnarray}
\lim_{x_* \to 0} f(x,y) = f_{_{\rm BB}}(x,y) \ ,
\label{bremscon7}
\end{eqnarray}
as required. In Figure~5 we use equation~(\ref{bremscon6}) to plot
$f(x,y)$ for several values of the low-energy cutoff $x_*$. We also
depict the equilibrium Wien spectrum $\wien(x) = (I_2/2) \, e^{-x}$,
with $I_2 = \Gamma(0,x_*)$. As expected, the solutions with nonzero
values of $x_*$ are able to fully equilibrate to the Wien spectrum
because the number of low-energy photons is finite in these cases.
Equilibration occurs more rapidly as $x_*$ is increased. Also
included for comparison is the solution $f_{_{\rm BB}}(x,y)$
corresponding to $x_*=0$, with an infinite number of photons
(eq.~[\ref{bremscon2}]). In this case, the spectrum is unable to
equilibrate. Our new solution for $f(x,y)$, given by
equation~(\ref{bremscon6}), therefore displays a more reasonable
physical behavior, and consequently it provides a rigorous basis for
models of the X-ray spectral evolution during AGN flares. In \S~5 we
discuss the results obtained when equation~(\ref{bremscon6}) is coupled
with a spatial transport model.

\begin{figure}
\hspace{15mm}
\includegraphics[width=125mm]{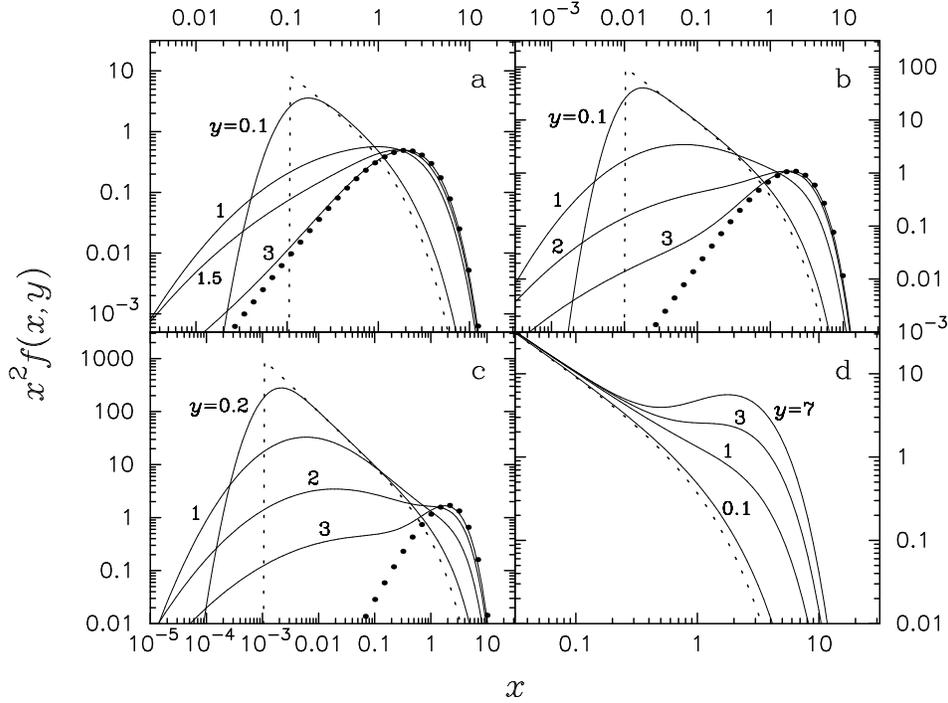}
\caption{Time-dependent photon energy distribution $x^2 f(x,y)$
resulting from the Comptonization of the modified bremsstrahlung initial
spectrum (eq.~[\ref{bremscon6}]; solid lines) plotted as a function of
$x$ for the indicated values of the dimensionless time $y$. Also
included for comparison are the corresponding initial spectra (eq.~[\ref{bremscon3}];
dotted lines) and the equilibrium Wien spectra (filled circles). The
values for the low-energy cutoff in the initial spectrum are (a) $x_* =
10^{-1}$; (b) $x_* = 10^{-2}$; (c) $x_* = 10^{-3}$; (d) $x_* = 0$. In
panel (d), we have used equation~(\ref{bremscon2}) to evaluate $x^2 f$,
and the Wien spectrum has been omitted since it does not exist in this
case. The solutions in panels (a) - (c) show proper equilibration to the
Wien spectrum as $y$ increases, whereas the distribution in panel (d)
does not due to the infinite number of photons in the initial spectrum.}
\end{figure}

\section{MODELS INCLUDING PHOTON ESCAPE}

The Green's function solution obtained in \S~3 describes the evolution
of the photon energy distribution $\green(x,x_0,y)$, but it does not
address spatial issues such as the distribution of photons within the
scattering cloud. In order to obtain the Green's function for the
occupation number, $\greenoc(x,x_0,\vec r,\vec r_0,y)$, we must solve
the spatial diffusion equation~(\ref{eq5}) for the radiation number
density $\greendens(\vec r,\vec r_0,y)$, and combine this information
with $\green(x,x_0,y)$ using equation~(\ref{eq9}). However, in many
applications the detailed spatial distribution is not needed, and in
such cases one may simply integrate over the volume of the cloud to
describe the radiation spectrum as a function of energy and time only.
In this section we take the latter approach and incorporate spatial
effects by using a simple escape-probability formalism that serves to
illustrate the general concept. More complex models with realistic
geometries can easily be accommodated within our framework, provided the
electron distribution in the cloud is homogeneous, as we have assumed.

\subsection{Escape time distribution}

Let us suppose that $N_0$ photons are injected into the cloud
with an arbitrary spatial distribution at time $t=t_0$. Then
the probability that a randomly-selected photon still remains inside
the cloud at time $t \ge t_0$ is given by the escape time distribution
\begeq
P(t) = {N_r(t) \over N_0} \ ,
\label{eq14}
\fineq
where
\begeq
N_r(t) = \oint_V n_r(\vec r,t) \, d^3 \vec r
\label{eq14b}
\fineq
denotes the number of photons contained inside the cloud at time $t$,
which is computed by integrating the radiation number density $n_r$ over
the volume $V$ of the cloud. It follows that $P(t)$ decreases
monotonically as a function of time from unity at $t=t_0$ to zero as $t
\to \infty$ due to the escape of photons from the cloud.

In a homogeneous plasma, the energy and spatial distributions evolve in
a decoupled manner as demonstrated in \S~2.2. The probability that a
photon randomly selected from the $N_0$ photons injected at time $t=t_0$
will subsequently escape in the time interval between $t$ and $t + dt$
can be expressed as
\begeq
P(t) \, {dt \over t_{\rm esc}} = - {1 \over N_0} {dN_r \over dt}
\, dt \ ,
\label{esc1}
\fineq
where the escape probability distribution $P(t)$ is given by
equation~(\ref{eq14}), and
\begeq
t_{\rm esc}(t)
= - \left({1 \over P}{d P \over dt}\right)^{-1}
\label{esc2}
\fineq
represents the mean escape time for photons remaining in the cloud at
time $t$. Note that $P(t) \, dt/t_{\rm esc}$ is also equal to the
fraction of the initial photons escaping between times $t$ and $t + dt$.
It is important to emphasize that, in general, the mean escape time
$t_{\rm esc}$ varies as a function of time $t$ due to the evolving
nature of the photon distribution inside the cloud, which is a result
of spatial diffusion. This variation can occur even when the electron
distribution in the cloud is homogeneous, as assumed here. By virtue
of equation~(\ref{esc1}), $P(t)$ clearly satisfies the normalization
requirement
\begeq
\int_{t_0}^\infty P(t) \, {dt \over t_{\rm esc}} = 1 \ ,
\label{esc3}
\fineq
since $N_r(t_0)=N_0$. Once the radiation number density distribution
$n_r(\vec r,t)$ has been determined by solving the appropriate diffusion
equation, the escape probability $P(t)$ can be evaluated using
equations~(\ref{eq14}) and (\ref{eq14b}). The result obtained for
$n_r(\vec r,t)$ will naturally depend on the shape of the cloud and also
on the initial spatial distribution of the photons at time $t=t_0$.
Consequently, these factors will also influence the escape time
distribution $P(t)$. Interested readers can find further details in
Sunyaev \& Titarchuk (1980).

Let us suppose that $P(t)$ has been computed in a given situation, and
that the particular solution for the photon energy distribution $f(x,y)$
has also been determined by utilizing equation~(\ref{convolve}). In this
case, the spectrum of the radiation remaining in the plasma at time $t$
can be obtained by forming the product of $f$ with the probability
distribution $P$. Specifically, we find that the number of photons with
dimensionless energy between $x$ and $x + dx$ still residing within the
cloud at time $t$ is given by
\begeq
N_x(x,t) \, dx \equiv P(t) \ x^2 \, f(x,y) \, dx
\ ,
\label{numdist1}
\fineq
where $y(t)=(t-t_0)n_e \sig c (kT_e/m_ec^2)$ (see eq.~[\ref{eq11}]).
Integration of $N_x$ with respect to energy yields the total number
of photons in the cloud,
\begeq
N_r(t) = \int_0^\infty N_x(x,t) \, dx \ .
\label{numdist2}
\fineq
The number of photons with energy between $x$ and $x
+ dx$ {\it escaping} from the cloud in the time interval
between $t$ and $t + dt$ is then given by
\begeq
\dot N_x(x,t) \, dt \, dx
\equiv t_{\rm esc}^{-1} \ N_x(x,t) \, dt \, dx
= t_{\rm esc}^{-1}
\ P(t) \, dt \ x^2 \, f(x,y) \, dx \ .
\label{esc4}
\fineq
In \S~4 we obtained particular solutions $f(x,y)$ corresponding to
several specific initial spectra. The associated results for the
escaping radiation spectrum $\dot N_x$ are discussed below.

\subsection{Escaping spectrum for monoenergetic initial condition}

In order to demonstrate the effects of spatial transport without undue
complexity, we shall assume here that $t_{\rm esc} = \rm constant$. This
implies that the spatial distribution of the photon sources is proportional
to the first eigenfunction of the diffusion operator, as discussed by
Sunyaev \& Titarchuk (1980). The specific result obtained for the initial
density distribution in the spherical, homogeneous case with $t_{\rm esc}
= \rm constant$ is
\begeq
n_r(r,t) \bigg|_{t=t_0} =
{N_0 \, \lambda^2 \over \sin(\lambda R/\ell)
- (\lambda R/\ell) \cos(\lambda R/\ell)}
\ {\sin(\lambda \, r/\ell) \over 4 \pi \ell^2 \, r}
\ ,
\label{mono0}
\fineq
where $N_0$ is the total number of photons in the initial distribution,
$R$ is the radius of the cloud, $\ell=(n_e \sig)^{-1}$ is the
(constant) mean free path for electron scattering, and the eigenvalue
$\lambda$ is the smallest positive root of the equation
\begeq
\tan\left({\lambda R \over \ell}\right) = \left({\lambda R \over
\ell}\right) \left(1-{\tilde\alpha R \over \ell}\right)^{-1}
\ .
\label{mono0b}
\fineq
The quantity $\tilde\alpha$ introduced in equation~(\ref{mono0b}) is a
constant of order unity that depends on the precise manner in which the
flux boundary condition is specified at the surface of the cloud using
the Eddington approximation. For example, Rybicki \& Lightman (1979) set
$\tilde\alpha=3^{1/2}$ in their equation~(1.124), whereas Sunyaev \&
Titarchuk set $\tilde\alpha=3/2$ in their equation~(A.3).

As $t$ increases from the initial time $t_0$, photons escape from the
cloud due to spatial diffusion. The resulting time-dependent solution
for the radiation number density is given by
\begeq
n_r(r,t) = n_r(r,t_0) \, e^{-(t-t_0)/t_{\rm esc}} \ ,
\label{mono0c}
\fineq
where $n_r(r,t_0)$ is the initial distribution [eq.~(\ref{mono0})]
and
\begeq
t_{\rm esc} = {3 \, \ell \over \lambda^2 \, c} = \rm constant
\label{mono0d}
\fineq
denotes the mean escape time. The probability that a photon injected
at time $t_0$ still remains within the cloud at time $t$ is therefore
given by
\begeq
P(t) = e^{-(t-t_0)/t_{\rm esc}} \ .
\label{mono1}
\fineq
We can now use equation~(\ref{numdist1}) to express the number of
photons with dimensionless energy between $x$ and $x + dx$ residing
within the cloud at time $t$ as
\begeq
N_x(x,t) \, dx
= e^{-(t-t_0)/t_{\rm esc}} \ x^2 \, f(x,y) \, dx
\ ,
\label{mono2}
\fineq
where $y(t) = \tilde y \, (t-t_0)/t_{\rm esc}$, and
\begeq
\tilde y \equiv
t_{\rm esc} \, n_e \sig c \, {k T_e \over m_e c^2}
= {\rm Max}(\tau,\tau^2) \, {k T_e \over m_e c^2}
\label{mono3}
\fineq
represents the mean value of the Compton $y$-parameter experienced by
photons before escaping from a cloud with scattering optical thickness
$\tau=R/\ell$ (Rybicki \& Lightman 1979). When $\tilde y \ga 1$, significant
Comptonization occurs before most of the radiation has escaped. In the
limit of small $\tilde y$, little change in spectral shape occurs during
the transient. The number of photons with energy between $x$ and $x +
dx$ escaping from the cloud in the time interval between $t$ and $t +
dt$ can be written as (see eq.~[\ref{esc4}])
\begeq
\dot N_x(x,t) \, dt \, dx
= t_{\rm esc}^{-1}
\ e^{-(t-t_0)/t_{\rm esc}} \, dt \ x^2 \, f(x,y) \, dx \ .
\label{mono4}
\fineq
In Figure~6, we plot the escaping photon distribution $\dot N_x(x,t)$
evaluated using equation~(\ref{mono4}), with $f(x,y)$ set equal to the
Green's function solution given by equation~(\ref{eq34}). Results are
presented for two values of the initial energy $x_0$ and two values of
the mean Compton parameter $\tilde y$. The curves depicted in Figure~6
represent the spectra that would be observed outside the cloud as the
initial burst of monochromatic (e.g., iron line) radiation is
reprocessed by electron scattering, and gradually escapes from the
cloud. For small values of $t-t_0$, the escaping spectrum is narrow and
centered on the initial energy $x_0$ since little scattering has
occurred. At intermediate times, the spectrum is broadened and the
detailed shape depends on the value of the initial energy $x_0$. The
photon distribution is characterized by spectral hardening for $x_0
\la 3$ and by spectral softening for $x_0 \ga 3$. For large values
of $t-t_0$, the spectrum has a Wien shape with a low normalization
because most of the radiation has escaped. Equilibration to the Wien
spectrum occurs more rapidly as $x_0$ and/or $\tilde y$ are increased.

\begin{figure}
\hspace{15mm}
\includegraphics[width=125mm]{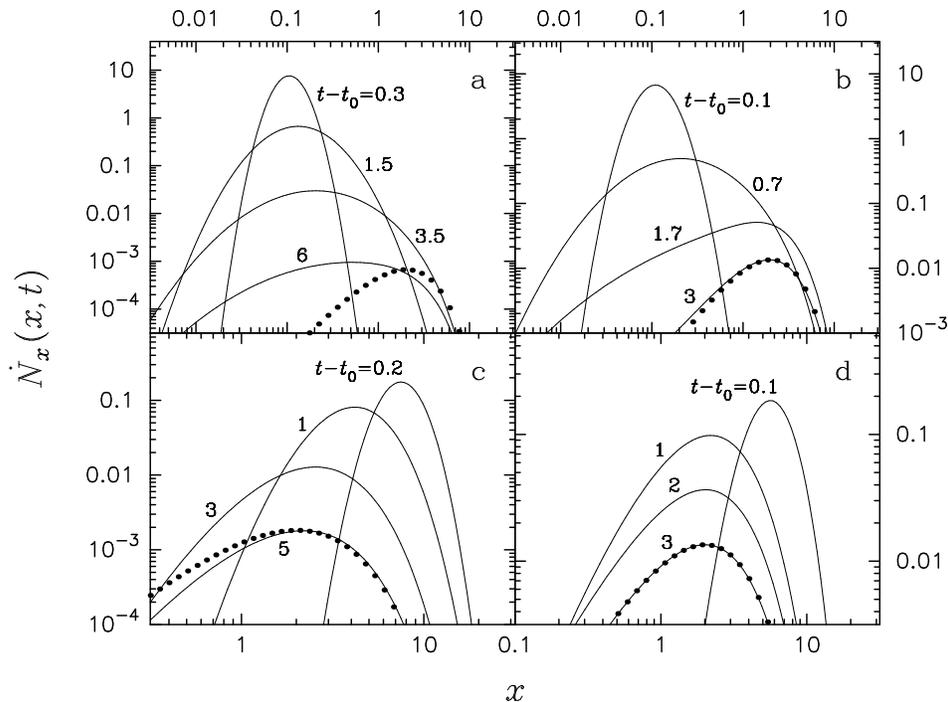}
\caption{Time-dependent escaping photon distribution $\dot N_x(x,t)$
(eq.~[\ref{mono4}]; solid lines) plotted in units of $t_{\rm esc}^{-1}$
for the case of a spectrum that is initially monoenergetic at time $t =
t_0$. The elapsed time $t-t_0$ in units of $t_{\rm esc}$ is indicated
for each curve, and the values of the initial photon energy $x_0$ and
the mean Compton parameter $\tilde y$ (eq.~[\ref{mono3}]) are given by
(a) $x_0 = 0.1$, $\tilde y = 0.2$; (b) $x_0 = 0.1$, $\tilde y = 1$; (c)
$x_0 = 10$, $\tilde y = 0.2$; and (d) $x_0 = 10$, $\tilde y = 1$. The
normalization of the curves decreases over time due to the loss of
radiation from the scattering cloud, as described by
equation~(\ref{mono1}). Included in each case is the Wien spectrum
(filled circles) evaluated at time $t_{\rm max}$, where $t_{\rm max}$ is
the maximum time for the sequence of curves in the panel.}
\end{figure}

\subsection{Escaping spectrum for bremsstrahlung initial condition}

It is also interesting to examine the distribution of radiation escaping
from the scattering cloud in the case of the modified bremsstrahlung
initial spectrum given by equation~(\ref{bremscon3}). We shall again
assume that $t_{\rm esc} = \rm constant$, and therefore we can employ
equation~(\ref{mono1}) for the escape probability distribution $P$. In
Figure~7, we plot the results for the escaping photon distribution $\dot
N_x(x,t)$ computed using equation~(\ref{mono4}), with $f(x,y)$ set equal
to the particular solution for the modified bremsstrahlung initial
spectrum given by equation~(\ref{bremscon6}). Results are presented for
two values of the low-energy cutoff $x_*$ and two values of the mean
Compton parameter $\tilde y$. For small values of $t-t_0$, the escaping
spectrum resembles the initial distribution. At intermediate times, the
spectrum displays the hardening that is characteristic of Comptonized
bremsstrahlung, and as $t$ increases, the spectrum approaches the Wien
distributution. The details of the spectral evolution depend on the
values of $x_*$ and $\tilde y$, which may allow the extraction of these
parameters from a sequence of spectra observed during a strong X-ray
transient. For example, let us suppose that the transient begins with
the impulsive injection into the corona of a pulse of soft
bremsstrahlung radiation with an initial spectrum given by
equation~(\ref{bremscon3}). As Figure~7 indicates, the shape of the
spectrum for energies $x \la 1$ at the time of appearance of the Wien peak,
combined with the overall amplitude of the spectrum at that time, is a
sensitive indicator of the unique values of $x_*$ and $\tilde y$. In
general, the low-energy spectrum increases with decreasing $x_*$ because
it takes longer for the additional soft photons to be upscattered, and
the overall amplitude of the spectrum at the time of appearance of the
Wien peak increases with increasing $\tilde y$ due to the competition
between Comptonization and photon escape. Indeed, the simple observation
of a Wien peak in the escaping spectrum would indicate that $\tilde y
\ga 1$, since otherwise the peak would occur after the normalization had
already declined by several orders of magnitude due to photon escape,
rendering the Wien bump unobservable. Increasing either $x_*$ or $\tilde
y$ results in more rapid equilibration to the Wien shape. The type of
spectral evolution depicted in Figure~7 is expected to occur when the
timescale for the impulsive injection of the bremsstrahlung radiation is
shorter that the timescale for the radiation to escape from the cloud.
In the next section we consider the situation with continual photon
injection, resulting in the evolution of the spectrum towards a
steady-state form.

\begin{figure}
\hspace{15mm}
\includegraphics[width=125mm]{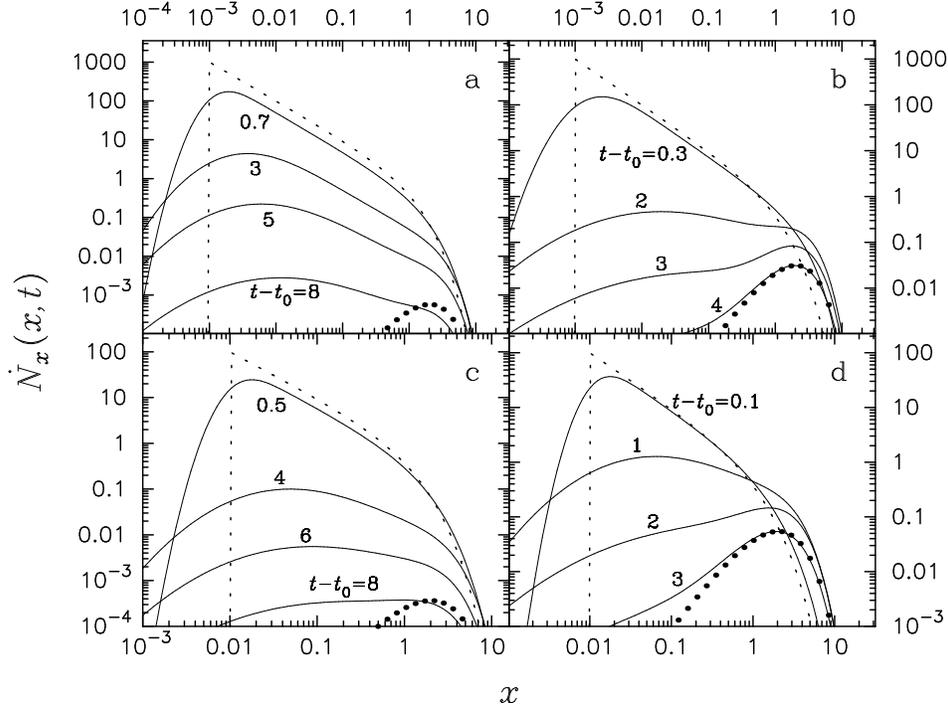}
\caption{Time-dependent escaping photon distribution $\dot N_x(x,t)$
(eq.~[\ref{mono4}]; solid lines) plotted in units of $t_{\rm esc}^{-1}$
for the case of the modified bremsstrahlung initial spectrum
(eq.~\ref{bremscon3}; dotted lines). The elapsed time $t-t_0$ in units
of $t_{\rm esc}$ is indicated for each curve, and the values of the
low-energy cutoff $x_*$ and the mean Compton parameter $\tilde y$ are
given by (a) $x_* = 10^{-3}$, $\tilde y = 0.2$; (b) $x_* = 10^{-3}$,
$\tilde y = 1$; (c) $x_* = 10^{-2}$, $\tilde y = 0.2$; and (d) $x_* =
10^{-2}$, $\tilde y = 1$. Included for comparison in each case is the
Wien spectrum (filled circles) evaluated at the time $t_{\rm max}$,
where $t_{\rm max}$ is the maximum time for the sequence of curves in
the panel.}
\end{figure}

\subsection{Generation of the steady-state Green's function}

The solution we have obtained for the Laplace transform $L(x,x_0,s)$ is
closely related to the steady-state photon energy distribution resulting
from the continual injection of monoenergetic radiation into a cloud of
scattering electrons that is leaking photons into the surrounding space.
In this section, we explore the connection in detail, and establish that
our time-dependent Green's function solution can be used to generate the
{\it steady-state} Green's function corresponding to a balance between photon
injection, thermal Comptonization, and photon escape. The steady-state
solution is one of the most important and widely applied spectral models
in X-ray data analysis, and it successfully accounts for the formation
of the power-law spectra commonly observed in the quiescent emission
from galactic and extragalactic sources.

Let us suppose that the monoenergetic photon source is turned on at time $t=0$.
In this situation, the number spectrum of the photons remaining in the cloud
at time $t$ can be written in terms of the Green's function as the convolution
\begeq
N_x(x,t) = \int_0^t \dot N_0 \, P(t) \,
x^2 \, \green(x,x_0,y) \, dt_0 \ ,
\label{steady2}
\fineq
where $\dot N_0$ expresses the rate of injection of fresh photons
with dimensionless energy $x_0$ into the cloud per unit time, and
$y(t) = n_e \sig c (k T_e/m_e c^2)(t-t_0)$ according to
equation~(\ref{eq11}). The number spectrum $N_x(x,t)$ is related
to the total number of photons in the cloud $N_r(t)$ via $N_r(t)
= \int_0^\infty N_x(x,t) \, dx$ (see eq.~[\ref{numdist2}]).
Transforming the variable of integration in equation~(\ref{steady2})
from $t_0$ to $\xi \equiv (t - t_0)/t_{\rm esc}$ and substituting
for $P$ using equation~(\ref{mono1}) yields
\begeq
N_x(x,t) = \int_0^{t/t_{\rm esc}}
t_{\rm esc} \, \dot N_0 \, e^{-\xi} \, x^2 \,
\green(x,x_0,\tilde y \, \xi) \, d \xi \ ,
\label{steady4}
\fineq
where $\tilde y = t_{\rm esc} \, n_e \sig c \, (k T_e/m_e c^2)$
(eq.~[\ref{mono3}]). The spectrum of the radiation escaping from
the cloud is given by
\begeq
\dot N_x(x,t) = \int_0^{t/t_{\rm esc}}
\dot N_0 \, e^{-\xi} \, x^2 \, \green(x,x_0,\tilde y \, \xi)
\, d \xi \ .
\label{steady4b}
\fineq

In the limit $t \to \infty$, a steady-state balance is obtained between
monoenergetic photon injection, thermal Comptonization, and photon escape.
The resulting steady-state spectrum of the escaping radiation, $\dot
N^{\rm std}_x(x)$, is given by
\begeq
\dot N^{\rm std}_x(x) = \lim_{t \to \infty} \dot N_x(x,t)
= \int_0^\infty \dot N_0 \,
e^{-\xi} \, x^2 \, \green(x,x_0,\tilde y \, \xi) \, d\xi \ .
\label{steady5}
\fineq
By construction, this represents the Green's function solution for the
time-independent problem. Transformation of the variable of integration from
$\xi$ to $y=\tilde y \, \xi$ now yields
\begeq
\dot N^{\rm std}_x(x) = {\dot N_0 \, x^2 \over \tilde y}
\int_0^\infty e^{-y/\tilde y} \, \green(x,x_0,y) \, dy \ .
\label{steady6}
\fineq
Comparing this expression with equation~(\ref{eq21}), we recognize that
the integral on the right-hand side is simply the Laplace transformation
of $\green(x,x_0,y)$ with the complex variable $s$ replaced by
$1/ \tilde y$. We can therefore immediately write
\begeq
\dot N^{\rm std}_x(x) = {\dot N_0 \, x^2 \over \tilde y}
\, L\left(x,x_0,{1 \over \tilde y}\right) \ .
\label{steady8}
\fineq
Substituting for $L(x,x_0,1/\tilde y)$ using equation~(\ref{eq23}) yields
the equivalent result
\begeq
\dot N^{\rm std}_x(x) = {\dot N_0 \over \tilde y}
\, {\Gamma(\tilde\mu-3/2) \over \Gamma(1+2\tilde\mu)} \,
x_0^{-2} \, e^{(x_0-x)/2} \,
M_{2, \, \tilde\mu}(\xmin) \ W_{2, \, \tilde\mu}(\xmax) \ ,
\label{steady9}
\fineq
where $\xmax$ and $\xmin$ are defined by equations~(\ref{eq24}),
and
\begeq
\tilde\mu \equiv \left({9 \over 4} + {1 \over \tilde y}
\right)^{1/2} \ .
\label{steady10}
\fineq
This is essentially the same solution to the stationary Kompaneets
equation derived by Sunyaev \& Titarchuk (1980) in their Appendix~B if
we set their separation constant $\gamma=1 / \tilde y$. The
normalization of our solution for $\dot N^{\rm std}_x(x)$ can be
confirmed by integrating equation~(\ref{steady9}) to show that the total
number of photons escaping from the cloud per unit time is given by
\begeq
\int_0^\infty \dot N^{\rm std}_x(x) \, dx = \dot N_0 \ ,
\label{steady12}
\fineq
as expected. Hence we have demonstrated that our solution for the
time-dependent Green's function can be used to generate the steady-state
Green's function. However, our solution also contains additional
information describing the evolution of the spectrum towards the steady
state, as discussed below.

\subsection{Evolution towards the steady-state spectrum}

We can use the results derived in \S~5.4 to study the evolution of the
photon distribution following the ``turning on'' of a continual source
of monochromatic radiation at time $t=0$. We assume that no photons are
initially present in the cloud. This scenario describes, for example, an
X-ray transient driven by the continual injection of iron line emission.
Whether or not a steady state is established depends on whether the
injection phase lasts longer than the characteristic timescale for the
escape of radiation from the cloud. In Figure~8, we plot a sequence of
curves describing the variation of the time-dependent escaping spectrum
$\dot N_x(x,t)$ computed using equation~(\ref{steady4b}), with
$\green(x,x_0,y)$ evaluated using equation~(\ref{eq34}). We also include
the steady-state spectrum $\dot N^{\rm std}_x(x)$ given by
equation~(\ref{steady9}). Two different values of the initial photon
energy $x_0$ and the mean Compton parameter $\tilde y$ are considered.
Note that as time increases, the spectrum broadens and the
time-dependent curves approach the corresponding steady-state solutions
as expected. The steady-state spectrum exhibits power-law wings around
$x = x_0$ when $\tilde y \la 1$ and $x_0 \la 1$ (see Rybicki \& Lightman
1979). For large values of $x$, the spectrum declines exponentially. The
final equilibrium represents a balance between Comptonization, photon
injection, and photon escape. The steady-state spectrum gives good
agreement with the quiescent X-ray spectra observed from many active
galaxies. However, with the availability of the new solution for the
time-dependent Green's function, it is now possible to model the {\it
transient} phase in the development of the steady-state spectrum using
equation~(\ref{steady4b}). This new theoretical framework improves our
ability to interpret spectral/temporal data associated with flares that
are driven by the injection of radiation on timescales shorter than that
required for the establishment of a steady state.

\begin{figure}
\hspace{15mm}
\includegraphics[width=125mm]{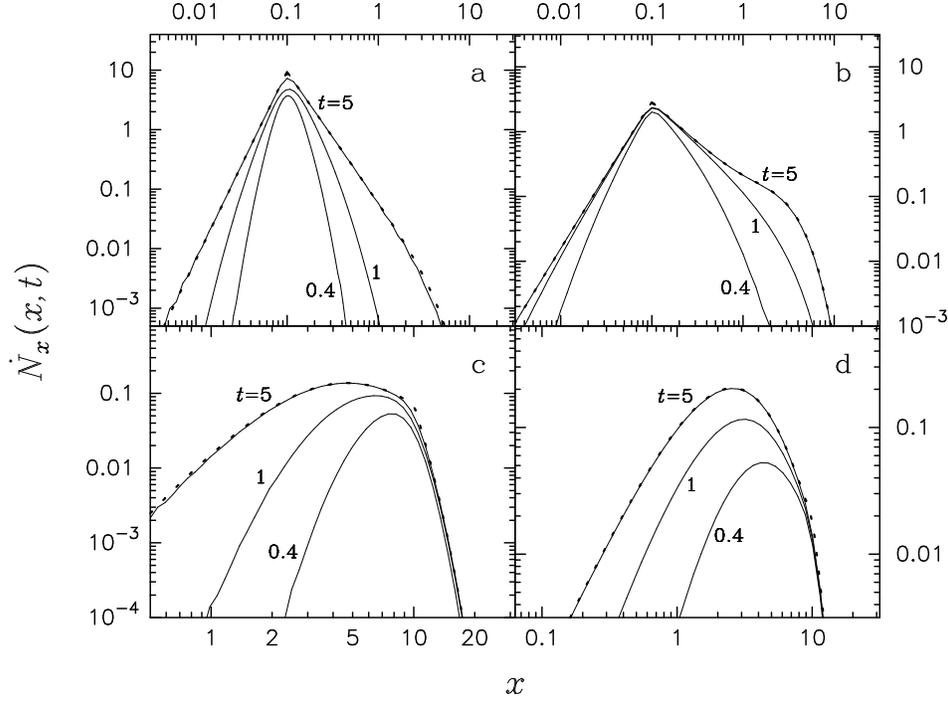}
\caption{Time-dependent escaping photon distribution $\dot N_x(x,t)$
(eq.~[\ref{steady4b}]; solid lines) plotted in units of the injection
rate $\dot N_0$ for the indicated values of the elapsed time $t$ in
units of $t_{\rm esc}$. The scenario considered here corresponds to the
continual injection of monochromatic radiation with energy $x_0$,
beginning at time $t=0$. The values of $x_0$ and the mean Compton parameter
$\tilde y$ are given by (a) $x_0 = 0.1$, $\tilde y = 0.2$; (b) $x_0 =
0.1$, $\tilde y = 1$; (c) $x_0 = 10$, $\tilde y = 0.2$; and (d) $x_0 =
10$, $\tilde y = 1$. As time increases, the spectrum broadens and the
time-dependent curves approach the steady-state solution $\dot N^{\rm
std}_x(x)$ (eq.~[\ref{steady9}]; dashed lines).}
\end{figure}

\section{CONCLUSIONS}

In this paper, we have derived several new results of importance in the
theory of time-dependent thermal Comptonization. The analysis is based on
the transport equation~(\ref{eq1}), which describes the effects of energy
diffusion (second-order Fermi energization), electron recoil, and spatial
diffusion. The spatial and energetic components of the problem were separated
by writing the Green's function for the occupation number $\greenoc$ as the
product of the energy Green's function $\green(x,x_0,y)$ and the spatial
distribution $\eta(\vec r,\vec r_0,y)$ (eq.~[\ref{eq9}]), which is valid
provided the scattering cloud has a steady, homogeneous structure. The
Green's function $\greenoc$ represents the response of the system to the
injection of a single photon with an arbitrary energy at an arbitrary time
and location within the cloud. The particular solution for the occupation
number distribution corresponding to a general source term can be obtained
by integrating $\greenoc$ using equation~(\ref{convolve1}).

The central result obtained in the subsequent analysis is the
closed-form solution for the Kompaneets Green's function,
$\green(x,x_0,y)$, given by equation~(\ref{eq34}) and plotted in
Figure~2. This fundamental expression describes the evolution of an
initially monoenergetic radiation spectrum in an infinite medium under
the influence of repeated Compton scattering. As such, it serves as the
``kernel'' for thermal Comptonization, from which one can obtain the
particular solution for any distribution of photon sources in time,
space, and energy. The validity of this new analytical solution was
confirmed in \S~3 via comparison with the results obtained for the
energy spectrum by integrating numerically the Kompaneets equation (see
Fig.~2). The availability of our closed-form solution for the Green's
function describing thermal Comptonization provides a valuable new tool
for the analysis and interpretation of variable X-ray spectra. In
particular, the Green's function represents an efficient alternative to
numerical integration of the Kompaneets equation that avoids the
complications and uncertainties associated with the imposition of
boundary conditions in the energy space.

The properties of the Green's function were explored in detail in \S~4,
where we confirmed that it displays proper equilibration to the Wien
spectrum at large times. We also reproduced Kompaneets' result
concerning the variation of the mean photon energy $\bar x$ as a
function of time (eq.~[\ref{moment10}]; Fig.~3), once a typographical
error made by Kompaneets is taken into account. Furthermore, we
obtained a new expression for the variation of the inverse-Compton
temperature associated with a monoenergetic initial spectrum (Fig.~4),
given by $\Tic/T_e=(1/4) \, \greenmom_4/\greenmom_3$, where the moments
$\greenmom_3$ and $\greenmom_4$ are evaluated using
equations~(\ref{moment10}) and (\ref{IC5}), respectively. We also used
our solution for the Green's function to derive a new, exact expression
for the spectrum resulting from the time-dependent Comptonization of a
bremsstrahlung initial spectrum, including a low-energy cutoff that
approximates the effect of self-absorption. The new solution
equilibrates to the Wien spectrum at large times (see
eq.~[\ref{bremscon6}] and Fig.~5).

Although the Green's function $\green(x,x_0,y)$ obtained here represents
the solution to the Kompaneets equation in an infinite medium, the
formalism we have developed can also incorporate the effects of spatial
transport in a homogeneous scattering cloud of any size and shape. In
\S~5, modifications associated with spatial transport in a cloud of
finite size were examined using a simple escape-probability model. In
this scenario, photons injected into the cloud with energy $x_0$ at some
initial time $t = t_0$ are subject to thermal Comptonization as they
diffuse out of the cloud. The analytical result for the escaping
radiation spectrum $\dot N_x(x,t)$ given by equation~(\ref{mono4}) was
plotted for monoenergetic and bremsstralung initial spectra in Figures~6
and 7, respectively. We also analyzed the case of the {\it continual}
injection of monoenergetic photons into a ``leaky'' cloud, starting at
time $t = 0$. A new time-dependent solution describing the gradual
build-up of the escaping radiation spectrum was obtained
(eq.~[\ref{steady4b}]), and it was explicitly confirmed that in the
limit $t \to \infty$, the time-dependent spectrum approaches the
steady-state solution given by Sunyaev \& Titarchuk (1980), describing a
balance between photon injection, thermal Comptonization, and photon
escape (see eq.~[\ref{steady9}] and Fig.~8).

\subsection{Relation to previous solutions}

The new solution for the Green's function obtained in \S~3.2 describes
the fundamental physics of thermal Comptonization in a homogeneous
plasma with steady properties. It therefore unifies and extends several
of the previously known analytical solutions. For example, the new
solution generalizes the analytical solution obtained by Zeldovich \&
Sunyaev (1969) and Payne (1980) for the Comptonization of soft photons
with $x \ll 1$, which satisfies the equation
\begeq
{\partial f \over \partial y} = {1 \over x^2} {\partial \over
\partial x} \left(x^4 {\partial f \over \partial x} \right) \ .
\label{con1}
\fineq
This equation does not include the electron recoil term proportional to
$f$ that appears in the full Kompaneets equation~(\ref{eq16}), and
therefore it only treats the stochastic energization of the photons. The
exact solution to equation~(\ref{con1}), which we refer to as the
``soft-photon Green's function,'' can be obtained using the Laplace
inversion integral (eq.~[\ref{eq26}]) by working in the limit $x \ll 1$,
$x_0 \ll 1$ (Titarchuk 2003, private communication). Along the inversion
contour, ${\rm Re} \, s = \gamma > 0$, and therefore it follows that
${\rm Re} \, \mu > 0$ since $\mu=(s+9/4)^{1/2}$ according to
equation~(\ref{eq25}). The Whittaker functions $W_{2, \, \mu}(x)$ and
$M_{2, \, \mu}(x)$ appearing in equation~(\ref{eq23}) for the Laplace
transform $L(x,x_0,s)$ consequently have the leading behaviors (see
eqs.~[\ref{whit1}] and [\ref{whit2}])
\begeq
M_{2, \, \mu}(x) \cong e^{-x/2} \, x^{\mu + 1/2} \ ,
\ \ \ \ \ 
W_{2, \, \mu}(x) \cong {\Gamma(2 \mu)
\over \Gamma(\mu - 3/2)}e^{-x/2} \, x^{-\mu + 1/2} \ ,
\label{con1b}
\fineq
for small $x$. Utilizing these asymptotic expressions for the Whittaker
functions in equation~(\ref{eq23}), we find that the Laplace transform
of the soft-photon Green's function is given by
\begeq
L_{\rm soft}(x,x_0,s) = {(x_0 x)^{-3/2} \over 2 \mu}
\left({\xmin \over \xmax}\right)^\mu
\ ,
\label{con1c}
\fineq
where $\xmax$ and $\xmin$ are defined by equations~(\ref{eq24}).
By performing the inverse Laplace transformation of $L_{\rm soft}$
using equation~(\ref{eq26}), we obtain for the soft-photon Green's
function the solution
\begeq
\greensoft(x,x_0,y) = {(x_0 x)^{-3/2} \, e^{-9y/4}
\over 2 \, \sqrt{\pi y}} \, \exp\left[{-(\ln x - \ln x_0)^2
\over 4 y}\right] \ ,
\label{con2}
\fineq
which satisfies equation~(\ref{con1}), as well as the initial condition
(see eq.~[\ref{eq17}])
\begeq
\greensoft(x,x_0,y) \Big|_{y=0} = x_0^{-2} \, \delta(x-x_0)
\label{con3} \ .
\fineq
Note that due to the neglect of the recoil effect, $\greensoft$ fails to
equilibrate to the Wien distribution as $y \to \infty$. In Figure~9 we
compare $\greensoft$ with the general Green's function $\green$
(eq.~[\ref{eq34}]) for two values of the initial photon energy,
$x_0=0.1$ and $x_0=1$. At early times (i.e., small values of $y$), the
two solutions agree closely. However, as $y$ increases, strong
disagreement becomes evident. In particular, at high energies (large
$x$), the $\greensoft$ distribution continues to increase far beyond the
saturation level indicated by the Wien spectrum. Conversely, our exact
solution for $\green$, which includes the effect of electron recoil,
equilibrates to the Wien form as required, and it therefore generalizes
the Zeldovich \& Sunyaev (1969) and Payne (1980) solution.

\begin{figure}
\hspace{15mm}
\includegraphics[width=125mm]{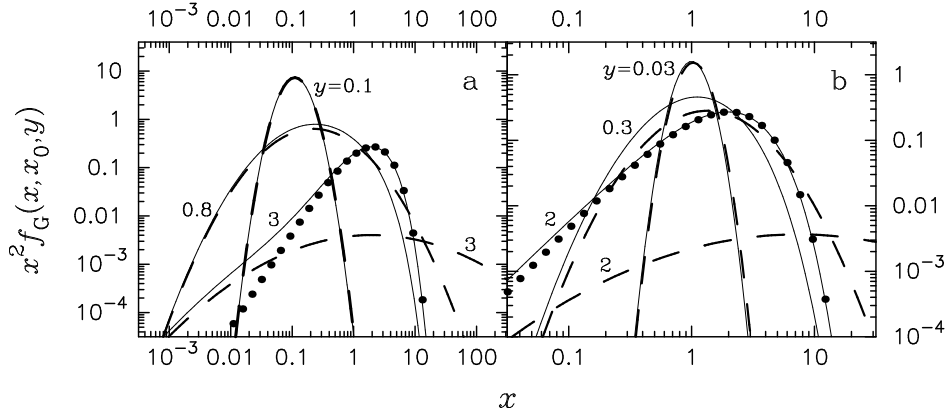}
\caption{Exact Green's function $x^2 \green(x,x_0,y)$ (eq.~[\ref{eq34}];
solid lines) plotted as a function of the dimensionless photon energy
$x$ for the indicated values of the dimensionless time $y$. The initial
photon energy is given by (a) $x_0 = 0.1$ and (b) $x_0 = 1$. Included
for comparison is the ``soft-photon Green's function,'' $x^2 \greensoft
(x,x_0,y)$ (eq.~[\ref{con2}]; dashed lines), which ignores the recoil
term in the Kompaneets equation~(\ref{eq16}), and therefore does not
equilibrate to the Wien spectrum $(1/2) \, x^2 e^{-x}$ (filled circles)
as $y \to \infty$. Conversely, $\green(x,x_0,y)$ satisfies the complete
Kompaneets equation, and therefore it approaches the Wien spectrum as
$y$ increases.}
\end{figure}

In \S~4.5 we considered the important problem of the Comptonization of a
``modified'' bremsstrahlung initial spectrum that includes a low-energy
cutoff at energy $x=x_*$, which simulates the effect of
self-absorption. Our new time-dependent analytical solution for the
resulting spectrum (eq.~[\ref{bremscon6}]) generalizes the analytical
solution obtained by Chapline \& Stevens (1973) and Becker \& Begelman
(1986), which sets the low-energy cutoff $x_* = 0$, and therefore
contains an infinite number of low-energy photons. While the $x_* = 0$
solution cannot equilibrate to the Wien spectrum, our new solution
satisfies this physical requirement, and it therefore represents a
distinct improvement over the previously known analytical result for
bremsstrahlung injection. We demonstrated in \S~4.5 that the low-energy
cutoff $x_*$ and the mean value of the Compton $y$-parameter, $\tilde
y$, can be estimated from observations of a sequence of spectra obtained
during a bremsstrahlung-driven X-ray flare. Knowledge of these
quantities yields constraints relating the electron temperature $T_e$
and optical thickness $\tau$ of the corona to the gas temperature $T$
and effective temperature $T_{\rm eff}$ in the source region (see
eqs.~[\ref{bremscon2c}] and [\ref{mono3}]).

The new solution for the Green's function obtained here also generalizes
the steady-state solution derived by Sunyaev \& Titarchuk (1980)
through the inclusion of the full time dependence of the spectral
evolution process. However, it should be pointed out that the Sunyaev \&
Titarchuk formalism is also applicable to the case of an inhomogeneous
scattering cloud, whereas the time-dependent solution obtained in
the present paper was developed under the assumption of a homogeneous
electron distribution. In future work we intend to investigate the
possibility of generalizing the time-dependent Green's function in
order to treat the inhomogeneous case as well. This would be a
significant improvement since observations suggest that the electron
distribution is probably inhomogeneous in a number of X-ray sources,
as discussed in \S~6.2. We briefly review the potential astrophysical
relevance of the results presented in this paper below.

\subsection{Astrophysical relevance}

The closed-form solution for the Green's function describing thermal
Comptonization (eq.~[\ref{eq34}]) provides a useful new tool for the
analysis and interpretation of variable X-ray spectra. Analytical
solutions offer far more physical insight than do numerical simulations,
and they also provide the flexibility to efficiently explore a large
region of parameter space. The new solution is relevant in a wide
variety of astrophysical situations involving thermal Comptonization,
such as studies of the reprocessing of the cosmological background
radiation (Shimon \& Rephaeli 2002; Colafrancesco et al. 1997), models
of thermal upscattering in gamma-ray bursts (Ghisellini \& Celotti 1999;
Liang et al. 1999), studies of variability and lags in the X-ray spectra
observed from AGNs and galactic black-hole candidates (Zdziarski et al.
2002; Lee et al. 2000), spectral reprocessing in X-ray bursts (Titarchuk
1988; Lapidus, Sunyaev, \& Titarchuk 1987), and the Comptonization of
line features in AGN spectra (Misra \& Kembhavi 1998; Wang et al. 1999).
The solution presented here is also perfectly suited for modeling the
acceleration of relativistic electrons in a turbulent plasma. In this
application, the terms proportional to $\partial \green / \partial x$
and $\green$ inside the parentheses on the right-hand side of
equation~(\ref{eq16}) describe the second-order Fermi acceleration of
the electrons and the effects of synchrotron/inverse-Compton losses,
respectively (Schlickeiser, Sievers, \& Thiemann 1987; Schlickeiser
1984; Lacombe 1977). The availability of our exact solution for the
Green's function therefore provides a unique opportunity to model
analytically the time-dependent development of the electron energy
spectrum during solar flares and other high-energy astrophysical
transients (e.g., Park, Petrosian, \& Schwartz 1997; Miller \& Ramaty
1989).

The best prospects for extragalactic, time-resolved X-ray spectroscopy
of accretion flows around black holes
are presented by nearby AGNs such as NGC 4051, NGC 4151, NGC 6814, and
MCG-6-30-15, which are among the brightest and most strongly variable
X-ray sources known. The rapid variability of these and other Seyfert
galaxies represents one of the most interesting puzzles in modern X-ray
astronomy. A large body of observational evidence suggests that as the
2-10$\,$keV flux increases in these sources, the X-ray spectrum becomes
softer (Lee et al. 2000; Merloni \& Fabian 2001; Nandra 2001). For
example, observations of the dwarf Seyfert nucleus of NGC 4395 reported
by Iwasawa et al. (2000) reveal rapid variability, including flares
during which the flux changes by a factor of 3-4 in a single half-day
{\it ASCA} observation. Doubling times as short as 100$\,$s have been
observed. The highly variable Seyfert galaxy NGC 6814 has displayed
factor of 3 changes in flux in 8 hours (K\"onig et al. 1997). The
characteristic softening observed during the Seyfert flares may reflect
inverse-Compton cooling of the corona due to the injection of copious
soft photons (possibly due to bremsstrahlung) from the inner region of
the accretion disk. The spectral evolution observed during these
transients may provide us with a direct opportunity to study
time-dependent Comptonization. However, it is also possible that the
variable X-ray spectra may be more correctly interpreted using a
sequence of steady-state models, if the timescale for the injection of
the photons exceeds the characteristic time for the photons to diffuse
out of the cloud (Sunyaev \& Titarchuk 1980).

The signature of time-dependent Comptonization is perhaps most
unambiguously observed in the complex cross-spectra produced by
combining the Fourier transformed lightcurves at two different X-ray
energies (van der Klis et al. 1987). The resulting structure of the
X-ray time lags contains detailed information on the ``Compton
reverberations'' taking place in the inner region of the disk/corona
(Payne 1980; Reynolds et al. 1999) that cannot be obtained using
standard spectral analysis (Kazanas, Hua, \& Titarchuk 1997).
Observations of hard X-ray time lags in AGNs and galactic black-hole
candidates are consistent with the predictions of thermal Comptonization
models (Zdziarski et al. 2002; Lee et al. 2000). The time lag is a general
consequence of the upscattering of soft photons by hot electrons (Payne
1980). It is worth emphasizing that the upscattering of soft
photons can take place even in situations where the electron temperature
is {\it decreasing} as a function of time due to inverse-Compton
cooling, as discussed in \S~6.3. Hence the spectral softening, combined
with the hard lags, strongly imply that thermal Comptonization is
playing a significant role in these sources. The hard lags are a
particularly important diagnostic, since they may provide an indelible
``imprint'' of the time-dependent process that can be detected even when
there is inadequate spectral-temporal resolution to accurately measure
the sequence of spectral curves.

The time-dependent nature of the analytical solution for the Green's
function obtained here provides a convenient basis for the theoretical
computation of the frequency-dependent lags. By exploiting the close
relationship between the Fourier and Laplace transforms, one can use the
exact expression for the Laplace transformation of the energy
distribution (eq.~[\ref{eq23}]) along with the Laplace transformation of
the escape probability distribution to obtain the Fourier transformation
of the escaping spectrum using the convolution theorem. This would yield
an analytical means for computing the complex cross-spectrum and
evaluating the X-ray time lags (van der Klis et al. 1987), which may
represent an efficient alternative to the Monte-Carlo based simulations
that are usually employed to compute the lags. The formalism developed
here focuses on a homogeneous scattering cloud, and there are some
indications that this may not be an appropriate model for certain
galactic black-hole candidates, in which the density in the corona may
vary as $1/r$ (B\"ottcher \& Liang 1998; Kazanas, Hua, \& Titarchuk
1997). However, this conclusion also depends on the shape of the cloud,
and it would therefore be interesting to extract the lags using the
analytical solution and compare them with the available X-ray data.
Moreover, it may be possible to extend the general analytical approach
taken here in order to treat inhomogeneous clouds. We plan to pursue
this possibility in future work.

In addition to the continuum emission discussed above, Seyfert galaxies
also appear to exhibit Fe K$\alpha$ line emission. This component of the
spectrum is quite broad, and in some cases highly variable (Wang, Zhou,
\& Wang 1999; Wang et al. 1999; Vaughan \& Edelson 2001). It is
currently not clear whether the width of the line features reflects
Doppler and gravitational broadening very close to the black hole
(Reynolds \& Wilms 2000), or perhaps broadening due to thermal
Comptonization in a surrounding corona (Misra \& Kembhavi 1998). Our
analytical solution for the Green's function may prove to be helpful in
resolving this issue since it provides an exact description of the
time-dependent reprocessing of the line emission in the corona. Sample
calculations illustrating the broadening of the line were presented in
Figure~6, where the effects of variations in the initial energy
$\epsilon_0$ relative to the electron thermal energy $kT_e$ were
explored, along with variations in the value of the mean Compton
parameter $\tilde y$. The analytical expression for the time-dependent
Green's function developed here is valid for arbitrary values of the
initial photon energy $\epsilon_0$, and therefore it can be used to
study the upscattering of an initial photon distribution with
$\epsilon_0 \ll kT_e$, or the downscattering of initial photons
with $\epsilon_0 \gg kT_e$. We reiterate that the previous analytical
solution for the time-dependent Green's function derived by Payne
(1980) and Zeldovich \& Sunyaev (1969) is valid only for the case of
very soft photons scattered by hot electrons, due to the neglect of
the recoil term in the Kompaneets equation.

\subsection{Thermal and dynamical effects}

It has long been recognized that the injection of large quantities of
soft radiation into a hot corona will almost certainly result in
substantial inverse-Compton cooling of the electrons (Payne 1980;
Lightman, Giacconi, \& Tananbaum 1978; Guilbert, Fabian, \& Ross 1982;
Becker \& Begelman 1986). This idea is further supported by the fact
that the spectral softening observed in Seyfert X-ray transients is
inconsistent with the hardening evident in the sequence of Comptonized
bremsstrahlung spectra plotted in Figure~7, based on the assumption of a
constant-temperature corona. We therefore conclude that the variation of
the electron temperature in the corona must be included in models for
the spectral evolution occurring during the flare. If the transient is
very intense,
the gas is expected to be radiation-dominated, and the electron
temperature should therefore track the inverse-Compton temperature of
the radiation (Becker \& Begelman 1986; Becker 1999). This issue has
been explored by B\"ottcher (2001), who performed Monte-Carlo
simulations of disk/corona models with temperatures that responded
self-consistently to the inverse-Compton cooling. They find that the
inclusion of thermal effects improved the agreement between the
predicted and observed X-ray time lags. If the cooling of the electrons
occurs on timescales that exceed the characteristic time for the photons
to diffuse out of the cloud, then the spectrum may be adequately
represented using a sequence of constant-temperature (though not
necessarily time-independent) models. On the other hand, if the electron
temperature varies on shorter timescales, then the entire problem of
thermal Comptonization must be reconsidered from first principles,
including a self-consistent treatment of the temperature. It is
currently unclear whether the temperature varies this rapidly during
typical Seyfert transients (see discussions in Payne 1980; Petrucci et
al. 2001; Nandra 2001; Malzac \& Jourdain 2000). We plan to explore this
question in future work.

Another issue of potential importance for models of the spectral
evolution during X-ray transients is the possible dynamical response of
the corona, driven by the heating and cooling of the gas. Bulk motions
may be important if the gas responds hydrodynamically to the energy
input, or if it is already moving, as in an accretion flow. Expansion of
the corona due to heating may deplete some of the energy from the
photons, and, conversely, contraction will lead to photon energization
via the first-order Fermi process (``bulk Comptonization''). Although we
have assumed here that the electron temperature $T_e$ and number density
$n_e$ remain constant during the transient, in general these two
quantities should be considered functions of time and position, since
both will respond to the hydrodynamics and thermodynamics of the gas.
Some progress has been made in understanding the effect of the bulk flow
on the radiation spectrum in idealized situations. For example, Colpi
(1988) has analyzed the effect of thermal Comptonization on photons
injected at an arbitrary radius into a freely-falling, spherical
accretion flow using a steady-state model. She finds that the
convergence of the flow can significantly alter the shape of the
escaping high-energy spectrum due to a combination of photon trapping
and Fermi energization. Similarly, Laurent \& Titarchuk (2001) find that
bulk Comptonization in quasi-spherical inflows can have a direct effect
on the steady-state power-law continuum emission produced. Following a
similar approach, it may be possible to incorporate the effects of bulk
motion within the framework of a time-dependent model such as the one
considered here.

In conclusion, we believe that the theoretical framework developed
in this paper will facilitate a more complete utilization of the
high-quality temporal and spectral data provided by current and future
orbital X-ray observatories. The new solution for the Green's function
generalizes and extends the previously known analytical solutions for
thermal Comptonization, and further extensions to cases involving
variable coronal temperatures and density structures may also be
possible. When combined with time-resolved spectra obtained during
bright X-ray flares in AGNs, these analytical solutions may allow the
determination of source parameters that previously were difficult to
interpret due to the necessity of running lengthy computer simulations.

The author is grateful to Demos Kazanas for several stimulating
conversations, and also to the referee, Lev Titarchuk, for providing
many useful comments which led to significant improvements in the manuscript.

\appendix

\section[]{Mathematical Details}

\subsection{Formal solution for the Laplace transform}

The Laplace transform
\begeq
L(x,x_0,s) \equiv \int_0^\infty e^{-s y} \, \green(x,x_0,y) \, dy
\label{aa1}
\fineq
of the Green's function distribution $\green(x,x_0,y)$ satisfies the
inhomogeneous ordinary differential equation
\begeq
{1 \over x^2} {d \over dx}
\left[x^4 \left(L + {dL \over dx}\right)\right]
- s \, L = - x_0^{-2} \, \delta(x-x_0) \ .
\label{aa2}
\fineq
The homogeneous equation obtained when $x \ne x_0$ has the solutions
\begeq
L(x,x_0,s) = \cases{
A \ x^{-2} \, e^{-x/2} \, M_{2, \, \mu}(x)
\ , & $x \le x_0$ \ , \cr
\phantom{stuff} & \phantom{stuff} \cr
B \ x^{-2} \, e^{-x/2} \, W_{2, \, \mu}(x)
\ , & $x \ge x_0$ \ , \cr
}
\label{aa3}
\fineq
where $M_{2, \, \mu}(x)$ and $W_{2, \, \mu}(x)$ denote Whittaker's
functions, and
\begeq
\mu(s) \equiv \left({9 \over 4} + s \right)^{1/2} \ .
\label{aa4}
\fineq
Asymptotic analysis indicates that these are the only functions that
yield convergent results for the photon number and energy densities. The
constants $A$ and $B$ appearing in equation~(\ref{aa3}) are determined
by imposing continuity and derivative jump conditions on the transform
$L$. The solution for $L$ must be continuous at $x = x_0$ in order to
avoid an infinite flux of photons in the energy space. The continuity
condition requires
\begeq
A \, M_{2, \, \mu}(x_0) = B \, W_{2, \, \mu}(x_0) \ .
\label{aa5}
\fineq
Furthermore, the derivative $dL/dx$ must display a jump at $x = x_0$
given by
\begeq
\lim_{\varepsilon \to 0} \ {dL \over dx} \Big|_{x_0+\varepsilon}
-{dL \over dx} \Big|_{x_0-\varepsilon} = - {1 \over x_0^4} \ ,
\label{aa6}
\fineq
which is obtained by integrating equation~(\ref{aa1}) in a narrow
range around $x = x_0$.

Utilization of the continuity and derivative jump conditions yields
for $A$ and $B$ the solutions
\begeq
A = {- e^{x_0/2} \, x_0^{-2} \, W_{2, \, \mu}(x_0) \over w(x_0)} \ ,
\ \ \ \ \ \ 
B = {- e^{x_0/2} \, x_0^{-2} \, M_{2, \, \mu}(x_0) \over w(x_0)} \ ,
\label{aa7}
\fineq
where the Wronskian of the two solutions is defined by
\begeq
w(x) \equiv M_{2, \, \mu}(x) \, {d \over dx} \, W_{2, \, \mu}(x)
- W_{2, \, \mu}(x) \, {d \over dx} \, M_{2, \, \mu}(x) \ .
\label{aa8}
\fineq
The Wronskian can be evaluated using equations~(13.1.22), (13.1.32),
and (13.1.33) of Abramowitz \& Stegun (1970), which yields
\begeq
w(x) = - \, {\Gamma(1+2\mu) \over \Gamma(\mu-3/2)} \ .
\label{aa9}
\fineq
Combining results, we find that the formal solution for the
Laplace transform is given by
\begeq
L(x,x_0,s) = {\Gamma(\mu-3/2) \over \Gamma(1+2\mu)} \,
x_0^{-2} \, x^{-2} \, e^{(x_0-x)/2} \,
\cases{
W_{2, \, \mu}(x_0) \, M_{2, \, \mu}(x) \ , & $x \le x_0$ \ , \cr
\phantom{stuff} \cr
M_{2, \, \mu}(x_0) \, W_{2, \, \mu}(x) \ , & $x \ge x_0$ \ , \cr
}
\label{aa10}
\fineq
or, equivalently,
\begeq
L(x,x_0,s) = {\Gamma(\mu-3/2) \over \Gamma(1+2\mu)} \,
x_0^{-2} \, x^{-2} \, e^{(x_0-x)/2} \,
M_{2, \, \mu}(\xmin) \ W_{2, \, \mu}(\xmax) \ ,
\label{aa11}
\fineq
where
\begeq
\xmin \equiv \min(x,x_0) \ , \ \ \ \ \ \ 
\xmax \equiv \max(x,x_0) \ .
\label{aa12}
\fineq
The solution for the Laplace transform $L(x,x_0,s)$ is closely related
to the Green's function for the steady-state problem developed by Sunyaev
\& Titarchuk (1980), as discussed in \S~5.4. Simple poles are located
where the function $\Gamma(\mu-3/2)$ diverges, which occurs when the
quantity $\mu-3/2$ is zero or a negative integer. The evaluation of
the residues at the simple poles is carried out in the next section,
and the inverse Laplace transformation is discussed in \S~3.2.

\subsection{Evaluation of the residues}

We found in \S~3 that the Green's function energy distribution can be
expressed as
\begeq
\green(x,x_0,y) =
- {1 \over 2\pi i} \int_N^O e^{sy} \, L(x,x_0,s) \, ds
- {1 \over 2\pi i} \int_P^Q e^{sy} \, L(x,x_0,s) \, ds
+ \sum_{n=1}^2 \, {\rm Res}(s_n) \ ,
\label{ac1}
\fineq
in the limit $r_1 \to \infty$, $r_2 \to 0$ (see Fig.~1), where the
transform $L(x,x_0,s)$ is given by equation~(\ref{aa10}) and ${\rm Res}
(s_n)$ is the residue associated with the pole at $s = s_n$. In our
problem, simple poles are located at $s_1=0$ and $s_2=-2$, and the
corresponding residues are computed using the formulas (Butkov 1968)
\begeq
{\rm Res}(s_1) = \lim_{s \to 0} \ s \, L(x,x_0,s) \, e^{sy} \ ,
\ \ \ \ \ \ \ 
{\rm Res}(s_2) = \lim_{s \to -2} \ (s + 2) \, L(x,x_0,s) \, e^{sy} \ .
\label{ac2}
\fineq
Because the poles correspond to the singularities of the function
$\Gamma(\mu-3/2)$, evaluation of the residues requires utilization
of the asymptotic results
\begeq
\lim_{s \to 0} \ s \, \Gamma(\mu-3/2) = 3 \ ,
\ \ \ \ \ \ \ 
\lim_{s \to -2} \ (s + 2) \, \Gamma(\mu-3/2) = -1 \ ,
\label{ac3}
\fineq
which follow from the definition of $\mu(s)$ given by
equation~(\ref{aa4}). We can show that the functions $M_{2, \, \mu}(x)$
and $W_{2, \, \mu}(x)$ reduce to combinations of Laguerre polynomials
and exponentials at the points $s=0$ and $s=-2$. In particular, using
equations~(13.1.2), (13.1.3), (13.1.32), and (13.1.33) from Abramowitz
\& Stegun (1970), we obtain for $s=0$ the results
\begeq
M_{2, \, \mu}(x) \bigg|_{s=0} = e^{-x/2} \, x^2 \ , \ \ \ \ \ \ \ \ 
W_{2, \, \mu}(x) \bigg|_{s=0} = e^{-x/2} \, x^2 \ .
\label{ac4}
\fineq
Likewise, for $s=-2$ we find that
\begeq
M_{2, \, \mu}(x) \bigg|_{s=-2} = e^{-x/2} \, x \, \left(1 - {x \over 2}
\right)\ , \ \ \ \ \ \ \ \ 
W_{2, \, \mu}(x) \bigg|_{s=-2} = - e^{-x/2} \, x \, (2 - x) \ .
\label{ac5}
\fineq
Combining equation~(\ref{aa10}) with equations~(\ref{ac2}) - (\ref{ac5}),
we establish that the residues are given by
\begeq
{\rm Res}(s_1) = {1 \over 2} \, e^{-x} \ , \ \ \ \ \ \ 
{\rm Res}(s_2) = {1 \over 2} \, e^{-x-2y} \ (2 - x) \, (2 - x_0)
\ x^{-1} x_0^{-1} \ .
\label{ac6}
\fineq

\subsection{Integration along the branch cut}

The remaining integrations in equation~(\ref{ac1}) along the upper and
lower segments $NO$ and $PQ$, respectively, need to be handled carefully
because of the presence of the branch cut extending from $s=-9/4$ to
$s=-\infty$ (see Fig.~1) in the definition of $\mu(s)$ (eq.~[\ref{aa4}]).
We proceed by transforming to a new variable of integration, $u$, defined
by
\begeq
u^2 \equiv - {9 \over 4} - s \ ,
\label{ac7}
\fineq
with $u=0$ at the branch point. This is useful because the
integration bounds in equation~(\ref{ac1}) now transform to the
convenient values $(0,\infty)$. In terms of $u$, equation~(\ref{aa4})
for $\mu(s)$ can be rewritten along the upper and lower segments as
\begeq
\mu(u) = i \, u \ , \ \ \ \ {\rm ~~segment}~NO \ ,
\label{ac8}
\fineq
\begeq
\mu(u) = - i \, u \ , \ \ \ \ {\rm segment}~PQ \ .
\label{ac9}
\fineq
The sign change between the two expressions for $\mu(u)$ along the two
segments reflect the two different branches of the complex square root
function. Using equation~(\ref{aa11}) to substitute for $L(x,x_0,s)$ in
equation~(\ref{ac1}) and transforming the variable of integration
from $s$ to $u$, with $ds = -2 \, u \, du$, we obtain
\begin{eqnarray}
\green(x,x_0,y) =
{e^{-9y/4} e^{(x_0-x)/2} \over \pi \, i \, x_0^2 \, x^2} \int_0^\infty
e^{-u^2 y} \, \bigg[{\Gamma(-3/2-i u) \over \Gamma(1-2iu)} \,
M_{2, \, -iu}(\xmin) \, W_{2, \, -iu}(\xmax) \phantom{SPAAACE}
\nonumber \\
\ - \ {\Gamma(-3/2+i u) \over \Gamma(1+2iu)} \,
M_{2, \, iu}(\xmin) \, W_{2, \, iu}(\xmax)
\bigg]
u \, du
+ \sum_{n=1}^2 \, {\rm Res}(s_n) \ , \phantom{SP}
\label{ac10}
\end{eqnarray}
where the residues are given by equations~(\ref{ac6}). From the definition
of the $W_{2, \, \mu}(x)$ function,
\begeq
W_{2, \, \mu}(x) \equiv {\Gamma(-2\mu) \over \Gamma(-\mu-3/2)}
\, M_{2, \, \mu}(x) + {\Gamma(2\mu) \over \Gamma(\mu-3/2)}
\, M_{2, \, -\mu}(x) \ ,
\label{ac11}
\fineq
it follows that
\begeq
W_{2, \, iu}(x) = W_{2, \, -iu}(x)
\ ,
\label{ac12}
\fineq
and therefore we can rewrite our expression for the Green's function as
\begin{eqnarray}
\green(x,x_0,y) =
{e^{-9y/4} e^{(x_0-x)/2} \over \pi \, i \, x_0^2 \, x^2} \int_0^\infty
e^{-u^2 y} \, W_{2, \, iu}(\xmax) \bigg[{\Gamma(-3/2-i u) \over
\Gamma(1-2iu)} \, M_{2, \, -iu}(\xmin) \phantom{SPAAACE}
\nonumber \\
\ - \ {\Gamma(-3/2+i u) \over \Gamma(1+2iu)} \,
M_{2, \, iu}(\xmin)
\bigg]
u \, du
+ \sum_{n=1}^2 \, {\rm Res}(s_n) \ . \phantom{SP}
\label{ac13}
\end{eqnarray}
Based on equation~(\ref{ac11}), we can show that the factor in square
brackets in equation~(\ref{ac13}) is given by
\begeq
\left[
{\Gamma(-3/2-i u) \over
\Gamma(1-2iu)} \, M_{2, \, -iu}(\xmin)
- {\Gamma(-3/2+i u) \over \Gamma(1+2iu)} \,
M_{2, \, iu}(\xmin)
\right]
= \ {\Gamma(-3/2-iu) \, \Gamma(-3/2+iu) \over \Gamma(2iu) \, \Gamma(1-2iu)}
W_{2, \, iu}(\xmin) \ .
\label{ac14}
\fineq
The products of gamma functions on the right-hand side of this
expression can be simplified using the identities
\begeq
\Gamma(2iu) \, \Gamma(1-2iu)
= {\pi \over 2 \, i \, \sinh(\pi u) \, \cosh(\pi u)} \ ,
\label{ac15a}
\fineq
\begeq
\Gamma(-3/2-iu) \, \Gamma(-3/2+iu)
= {16 \, \pi \over \cosh(\pi u) \, (1 + 4 u^2) \, (9 + 4 u^2)}
\ .
\label{ac15b}
\fineq
Combining equations~(\ref{ac6}), (\ref{ac13}), (\ref{ac14}),
(\ref{ac15a}), and (\ref{ac15b}), we obtain after some algebra
\begin{eqnarray}
\green(x,x_0,y) =
{32 \over \pi} \ e^{-9y/4} x_0^{-2} x^{-2} e^{(x_0-x)/2}
\int_0^\infty e^{-u^2 y} \, {u \, \sinh(\pi u) \over
(1 + 4 u^2)(9 + 4 u^2)} \phantom{SPAAAAAACE} \nonumber \\
\times \ W_{2, \, i u}(x_0) \,
W_{2, \, i u}(x) \, du
\ + \ {e^{-x} \over 2}
\ + \ {e^{-x-2y} \over 2} \ {(2 - x) \, (2 - x_0) \over
\ x_0 \, x} \ .
\label{ac16}
\end{eqnarray}
This is our final result for the Green's function, which appears as
equation~(\ref{eq34}) in the main text. Series expansions for the
Whittaker functions are provided in \S~A.4. A self-contained FORTRAN
program that evaluates the Whittaker functions and performs the
integration in equation~(\ref{ac16}) to yield the Green's function is
available from the author upon request. We point out that a related
integral involving a single Whittaker function was worked out by
Kompaneets (1957) in collaboration with I. M. Gelfand in order to
describe the variation of the mean photon energy due to thermal
Comptonization, as discussed in \S~4.2 (see eq.~[\ref{moment10}]).

\subsection{Series expansions for the Whittaker functions}

In order to carry out the integration in the analytical solution for
the Green's function given by equations~(\ref{eq34}) and (\ref{ac16}),
we need to be able to evaluate the Whittaker function $W_{2, \, i u}(x)$
for both large and small values of the argument $x$. Furthermore, the
solution we have obtained for the optically thin bremsstrahlung initial
spectrum (eq.~[\ref{bremscon6}]) requires the evaluation of the functions
$W_{1, \, i u}(x_*)$, $W_{0, \, i u}(x_*)$, $W_{-1, \, i u}(x_*)$, and
$W_{-2, \, i u}(x_*)$ for small values of the low-energy cutoff
$x_*$ (see eq.~[\ref{bremscon3}]). We shall develop the necessary
expansions below.

\subsubsection{Expansion for Small $x$}

Based on the definition of $W_{\kappa, \, \mu}(x)$ given by
equation~(13.1.34) of Abramowitz \& Stegun (1970), we can write
\begeq
W_{\kappa, \, i u}(x) \equiv {\Gamma(-2iu) \over \Gamma(1/2-\kappa-iu)}
\, M_{\kappa, \, i u}(x) + {\Gamma(2iu) \over \Gamma(1/2-\kappa+iu)}
\, M_{\kappa, \, - i u}(x) \ .
\label{whitser1}
\fineq
The two terms on the right-hand side are complex conjugates of each other,
and therefore $W_{\kappa, \, i u}(x)$ is purely real. Hence we obtain
\begeq
W_{\kappa, \, i u}(x) = 2 \, {\rm Re}\left\{{\Gamma(-2iu) \over
\Gamma(1/2-\kappa-iu)} \, M_{\kappa, \, i u}(x)\right\} \ .
\label{whitser2}
\fineq
For small values of $x$, the function $M_{\kappa, \, \mu}(x)$ can be
evaluated using the series
\begeq
M_{\kappa, \, i u}(x) = e^{-x/2} \, x^{iu + 1/2} \, \sum_{n=0}^\infty
{(1/2-\kappa+iu)_n \over (1 + 2iu)_n} \, {x^n \over n!} \ ,
\label{whitser3}
\fineq
where $(a)_n$ denotes the Pochhammer symbol, defined by (Abramowitz
\& Stegun 1970)
\begeq
(a)_n \equiv \cases{
1 \ , & $n=0$ \ , \cr
a (a+1) \cdots (a+n-1) \ , & $n \ge 1$ \ . \cr
}
\label{whitser4}
\fineq
Combining expressions, we obtain finally
\begeq
W_{\kappa, \, i u}(x) = 2 \, e^{-x/2} \, x^{1/2}
\, \sum_{n=0}^\infty {x^n \over n!} \
{\rm Re}\left\{
{\Gamma(-2iu) \, x^{i u} \over \Gamma(1/2-\kappa-iu)}
{(1/2-\kappa+iu)_n \over (1+2iu)_n} \right\}
\ .
\label{whitser5}
\fineq
This expansion can be used to evaluate each of the Whittaker
functions $W_{1, \, i u}(x_*)$, $W_{0, \, i u}(x_*)$,
$W_{-1, \, i u}(x_*)$, and $W_{-2, \, i u}(x_*)$ appearing in
equation~(\ref{bremscon6}).

An application of special interest is the function $W_{2, \, i u}(x)$
appearing in equation~(\ref{eq34}) for the Green's function. We obtain
in this case
\begeq
W_{2, \, i u}(x) = 2 \, e^{-x/2} \, x^{1/2}
\, \sum_{n=0}^\infty {x^n \over n!} \
{\rm Re}\left\{
{\Gamma(-2iu) \, x^{i u} \over \Gamma(-3/2-iu)}
{(-3/2+iu)_n \over (1+2iu)_n} \right\}
\ .
\label{whitser6}
\fineq
In practice, we use this expansion to evaluate $W_{2, \, i u}(x)$
for $x \le x_{\rm mid}(u)$, where
\begeq
x_{\rm mid}(u) \equiv \cases{
20 \ , & $u \le 10/3$ \ , \cr
10 + 3 \, u \ , & $u \ge 10/3$ \ . \cr
}
\label{whitser7}
\fineq
We emphasize that the functions $W_{\kappa, \, i u}(x)$ and
$W_{2, \, i u}(x)$ are purely real, despite the appearance
of the imaginary number $i$ in equations~(\ref{whitser5}) and
(\ref{whitser6}). The complex gamma function $\Gamma(z)$ can
be evaluated using equation~(6.1.48) from Abramowitz \& Stegun
(1970).

\subsubsection{Expansion for Large $x$}

The Whittaker functions $W_{1, \, i u}(x_*)$, $W_{0, \, i u}(x_*)$,
$W_{-1, \, i u}(x_*)$, and $W_{-2, \, i u}(x_*)$ appearing in
equation~(\ref{bremscon6}) for the bremsstrahlung particular solution
need only be evaluated for small values of the low-energy cutoff $x_*$,
and therefore equation~(\ref{whitser5}) is sufficient for this case.
However, the function $W_{2, \, i u}(x)$ appearing in equation~(\ref{eq34})
for the Green's function must be evaluated for general, positive values of
$x$. We can use equation~(\ref{whitser6}) to compute $W_{2, \, i u}(x)$ when
$x \le x_{\rm mid}(u)$. For larger values of $x$, a convergent series for
$W_{2, \, i u}(x)$ can be obtained by combining equations~(13.1.33) and
(13.5.2) of Abramowitz \& Stegun (1970), which yields
\begeq
W_{2, \, i u}(x) = e^{-x/2} \, x^2 \,
\sum_{n=0}^\infty
{(-3/2 + i u)_n \, (-3/2 - i u)_n \over n! \, (-x)^n}  \ .
\label{whitser8}
\fineq
The Pochhammer symbols in this equation are complex conjugates
of each other, and therefore our result for $W_{2, \, i u}(x)$ can
be rewritten as the completely real expression
\begeq
W_{2, \, i u}(x) = e^{-x/2} \, x^2 \, \left\{
1 + \sum_{n=1}^\infty {(-x)^{-n} \over n!} \,
\prod_{m=1}^n \, \left[\left({5 \over 2} - m \right)^2 + u^2
\right] \right\} \ .
\label{whitser9}
\fineq
We use this formula to evaluate $W_{2, \, i u}(x)$ when
$x > x_{\rm mid}(u)$, where $x_{\rm mid}(u)$ is defined by
equation~(\ref{whitser7}). In performing the numerical integration
in equation~(\ref{eq34}), it is sufficient to treat the domain
$0 \le u \le 20$ because larger values of $u$ make a negligible
contribution to the integral.

\bsp

\label{lastpage}

\end{document}